\documentclass[conference]{IEEEtran}

\usepackage[utf8]{inputenc}
\usepackage{amsmath,amssymb,amsfonts}
\usepackage{graphicx}
\usepackage{textcomp}
\usepackage{xcolor}
\usepackage{booktabs} 
\usepackage{multirow} 
\usepackage{enumitem} 
\usepackage{xurl}     
\usepackage{hyperref} 
\usepackage{cite}     
\usepackage{listings} 
\usepackage{tabularx} 
\usepackage{microtype} 

\hypersetup{
    colorlinks=true,
    linkcolor=blue,
    filecolor=magenta,      
    urlcolor=blue, 
    citecolor=green,
    pdftitle={Agentic AI Identity and Access Management: A New Approach},
    pdfpagemode=FullScreen,
    breaklinks=true 
}

\usepackage{calc} 
\newlength{\figwidth}
\lstdefinestyle{customjson}{
    basicstyle=\ttfamily\footnotesize,
    breaklines=true,
    breakatwhitespace=true,
    columns=fullflexible, 
    postbreak=\mbox{\textcolor{red}{$\hookrightarrow$}\space},
    showstringspaces=false,
    commentstyle=\color{gray},
    keywordstyle=\color{blue},
    stringstyle=\color{purple},
    numbers=left,
    numberstyle=\tiny\color{gray},
    stepnumber=1,
    numbersep=5pt,
    backgroundcolor=\color{lightgray!10},
    frame=tb,
    tabsize=2,
    captionpos=b,
    escapeinside={\%*}{*)}
}

\newcommand{\orcidicon}[1]{%
    \href{https://orcid.org/#1}{%
        \includegraphics[width=10pt]{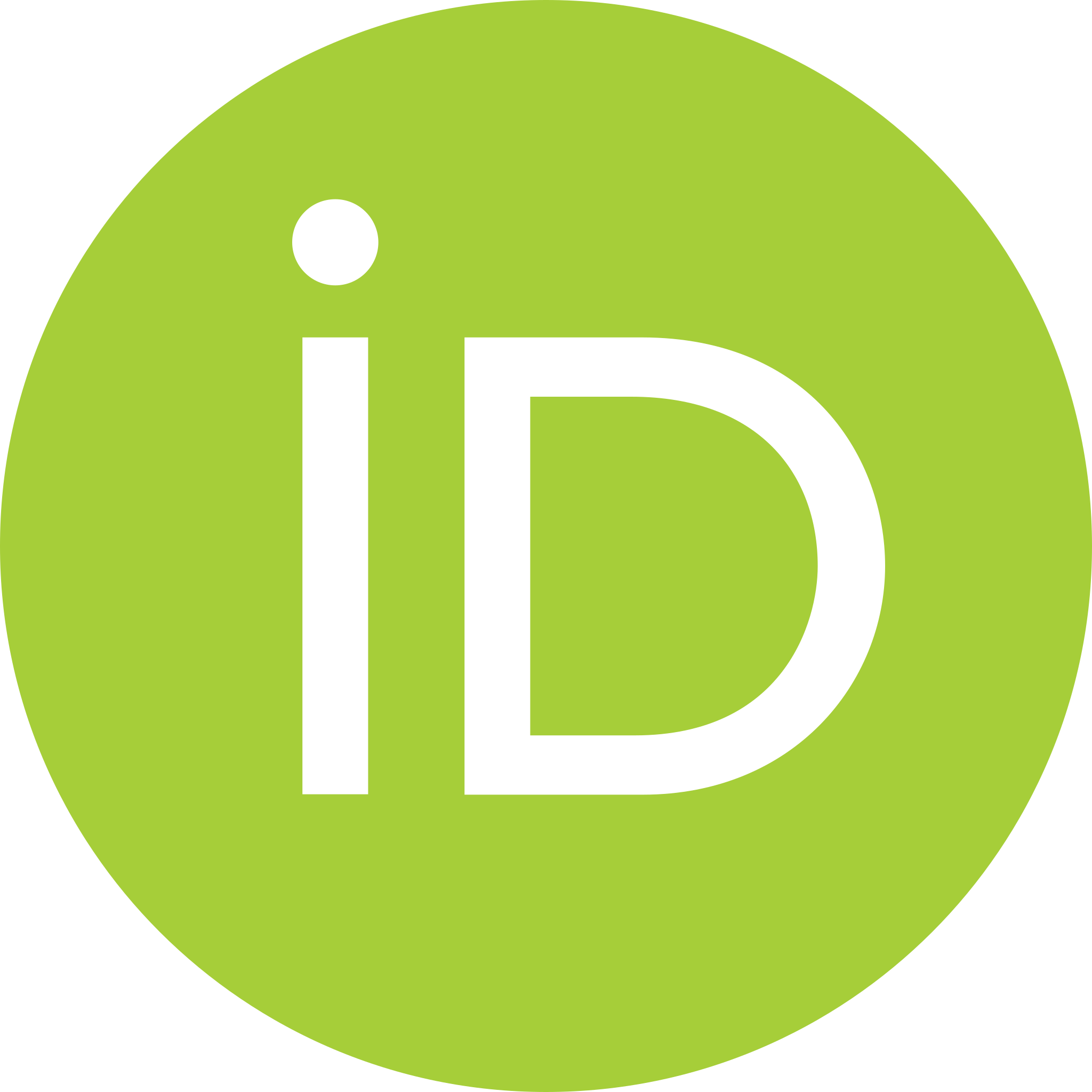}
    }%
}
\usepackage{academicons} 
\usepackage{xcolor} 

\lstdefinestyle{customcode}{
    basicstyle=\ttfamily\footnotesize,
    breaklines=true,
    breakatwhitespace=true,
    postbreak=\mbox{\textcolor{red}{$\hookrightarrow$}\space},
    showstringspaces=false,
    commentstyle=\color{gray},
    keywordstyle=\color{blue},
    stringstyle=\color{purple},
    numbers=left,
    numberstyle=\tiny\color{gray},
    stepnumber=1,
    numbersep=5pt,
    backgroundcolor=\color{lightgray!10},
    frame=tb,
    tabsize=2,
    captionpos=b,
    escapeinside={\%*}{*)},
    columns=flexible
}
\lstdefinestyle{customrego}{
    basicstyle=\ttfamily\footnotesize,
    morekeywords={package, default, allow, input, some, import}, 
    sensitive=true,
    breaklines=true,
    breakatwhitespace=true,
    columns=flexible, 
    postbreak=\mbox{\textcolor{red}{$\hookrightarrow$}\space},
    showstringspaces=false,
    commentstyle=\color{green!70!black}, 
    keywordstyle=\color{blue},
    stringstyle=\color{purple},
    numbers=left,
    numberstyle=\tiny\color{gray},
    stepnumber=1,
    numbersep=5pt,
    backgroundcolor=\color{lightgray!10},
    frame=tb,
    tabsize=2,
    captionpos=b,
    escapeinside={\%*}{*)}
}

\begin{document}

\title{A Novel Zero-Trust Identity Framework for Agentic AI: Decentralized Authentication and Fine-Grained Access Control}

\author{
     \IEEEauthorblockN{Ken Huang\IEEEauthorrefmark{1}\orcidicon{0009-0004-6502-3673}, Vineeth Sai Narajala\IEEEauthorrefmark{2}\orcidicon{0009-0007-4553-9930}, John Yeoh\IEEEauthorrefmark{3}, Json Ross\IEEEauthorrefmark{4}, Mahesh Lambe\IEEEauthorrefmark{5},}
     \IEEEauthorblockN{Ramesh Raskar\IEEEauthorrefmark{6}, Youssef Harkati\IEEEauthorrefmark{7}, Jerry Huang\IEEEauthorrefmark{8}, 
     Idan Habler \IEEEauthorrefmark{9}
     Chris Hughes\IEEEauthorrefmark{10}}
     
     \IEEEauthorblockA{\IEEEauthorrefmark{1}Fellow and Co-Chair, CSA AI Safety Working Groups}
     \IEEEauthorblockA{\IEEEauthorrefmark{2}Application Security Engineer, Amazon Web Services}
     \IEEEauthorblockA{\IEEEauthorrefmark{3}Chief Scientific Officer, EVP, Cloud Security Alliance}
     \IEEEauthorblockA{\IEEEauthorrefmark{4}Product Security Principal, Salesforce}
     \IEEEauthorblockA{\IEEEauthorrefmark{5}MIT NANDA Coauthor, Stanford GSB alumnus}
     \IEEEauthorblockA{\IEEEauthorrefmark{6}Professor, MIT; Founder, MIT Project Nanda for Agentic Web}
     \IEEEauthorblockA{\IEEEauthorrefmark{7}Co-founder and CTO, BrightOnLABS}
     \IEEEauthorblockA{\IEEEauthorrefmark{8}Researcher, The University of Chicago}
     \IEEEauthorblockA{\IEEEauthorrefmark{9}Independent Researcher},
     \IEEEauthorblockA{\IEEEauthorrefmark{10}Founder, Resilient Cyber}
 }

\maketitle

\begin{abstract}
Traditional Identity and Access Management (IAM) systems, primarily designed for human users or static machine identities via protocols such as OAuth, OpenID Connect (OIDC), and SAML, prove fundamentally inadequate for the dynamic, interdependent, and often ephemeral nature of AI agents operating at scale within Multi Agent Systems (MAS) – a computational system composed of multiple interacting intelligent agents that work collectively.

This paper posits the imperative for a novel Agentic AI - IAM framework: We deconstruct the limitations of existing protocols when applied to MAS, illustrating with concrete examples why their coarse-grained controls, single-entity focus, and lack of context-awareness falter. We then propose a comprehensive framework built upon rich, verifiable Agent Identities (IDs), leveraging Decentralized Identifiers (DIDs) and Verifiable Credentials (VCs), that encapsulate an agent's capabilities, provenance, behavioral scope, and security posture. 

Our framework includes an Agent Naming Service (ANS) for secure and capability-aware discovery, dynamic fine-grained access control mechanisms, and critically, a unified global session management and policy enforcement layer for real-time control and consistent revocation across heterogeneous agent communication protocols. We also explore how Zero-Knowledge Proofs (ZKPs) enable privacy-preserving attribute disclosure and verifiable policy compliance.

We outline the architecture, operational lifecycle, innovative contributions, and security considerations of this new IAM paradigm, aiming to establish the foundational trust, accountability, and security necessary for the burgeoning field of agentic AI and the complex ecosystems they will inhabit.
\end{abstract}


\section{Introduction}

Do we need a new approach for Agentic AI Identity Management? The failure to address the unique identity challenges posed by AI agents operating in Multi-Agent Systems (MAS) could lead to catastrophic security breaches, loss of accountability, and erosion of trust in these powerful technologies. For instance, without robust agent-specific IAM, a compromised autonomous agent in a financial system could cascade unauthorized transactions, or a swarm of interacting agents in critical infrastructure could be manipulated with devastating consequences. In this Cloud Security Alliance paper \cite{Huang2025aAgenticIAMApproach}, We listed initial reasons and approach. This paper expanded on our previous paper and proposed a more robust approach. (ABAC~\cite{Hu2019ABAC}, PBAC~\cite{Yaqub2025PBAC}, JIT~\cite{Shastri2025JIT}, \cite{Ferber1999MAS})

The core problem this current paper addresses is the fundamental mismatch between existing IAM paradigms (e.g., OAuth 2.0, OpenID Connect (OIDC), SAML) and the unique characteristics of AI agents in MAS. These agents exhibit autonomy, ephemerality, dynamically evolving capabilities, complex trust relationships, and operate at an unprecedented scale. Their actions carry direct consequences, demanding robust accountability. Delegated authority can cascade through multiple agents, obscuring responsibility if not managed appropriately. The European Union's AI Act \cite{EUAIAct2023} and similar regulatory initiatives underscore the growing societal demand for transparency, accountability, and human oversight in AI systems, making robust agent IAM an unavoidable prerequisite.

In a Wall Street Journal article, on May 17, 2025, Rosenbush \cite{Rosenbush2025WSJ} discusses the challenges AI agents encounter in accessing applications, APIs, and websites, emphasizing the need for new authentication methods beyond traditional human-centric options.

Inspired by preliminary discussions on IDs for AI systems \cite{Chan2024IDsAI}, this paper examines the limitations of current IAM protocols in MAS settings and illustrate, through concrete examples, how their coarse-grained permissions, single-entity assumptions, limited inclusion of Non-Human Identities (NHIs) and lack of contextual adaptability fall short and propose a need for a new, holistic Agentic AI IAM framework. We contend that merely adapting existing protocols is insufficient. Instead, a purpose-built approach is required, one that redefines agent identity, incorporates novel cryptographic primitives, and establishes new mechanisms for discovery, layered authentication, access control, and real-time policy enforcement tailored to the agentic paradigm.

This paper makes the following contributions:
\begin{itemize}
    \item It critically analyzes the inadequacies of traditional IAM protocols (OAuth, OIDC, SAML) in the context of MAS, providing concrete examples of their failure points.
    \item It defines the essential components of a rich, verifiable, and dynamic AI Agent Identity (ID), leveraging Decentralized Identifiers (DIDs) and Verifiable Credentials (VCs).
    \item It proposes a layered Agentic AI IAM architectural framework incorporating DIDs, VCs, Zero-Knowledge Proofs (ZKPs), an Agent Naming and Discovery Service (ANS) \cite{ans}, dynamic access control models, and a novel unified global session management and policy enforcement layer.
    \item It details how this framework addresses the lifecycle of agent IAM, from identity creation and attestation to runtime authorization, logging, monitoring, and incident response.
    \item It compares centralized, decentralized, and federated deployment models for this framework, offering guidance on their applicability, and analyzes security considerations using the MAESTRO framework.
\end{itemize}

The remainder of this paper is structured as follows: Section II elaborates on the imperative for a new agentic IAM paradigm by dissecting the limitations of traditional IAM. Section III defines the multifaceted nature of an AI agent's identity. Section IV presents the proposed Agentic AI IAM framework architecture. Section V discusses the operational use cases of Agent IDs within this framework. Section VI analyzes deployment models and governance. Section VII details security considerations. Section VIII highlights the innovative contributions. Section IX discusses future work, and Section X concludes.

\section{The Imperative for a New Agentic IAM Paradigm}

The rise of MAS necessitates a fundamental rethinking of how we manage identity and access. While traditional IAM protocols have served well for human-centric and simpler machine-to-machine interactions via service accounts, their core assumptions and mechanisms break down when faced with the complexities of autonomous, interacting AI agents.

\subsection{Revisiting Traditional IAM}

Protocols like OAuth 2.0 \cite{Hardt2012OAuth}, OpenID Connect (OIDC) \cite{OpenID2014OIDC}, and SAML \cite{OASIS2005SAML} are ubiquitous for authentication and authorization. Alongside these, foundational enterprise protocols such as Kerberos~\cite{Neuman2005Kerberos} for domain authentication and LDAP~\cite{Sermersheim2006LDAP} for directory services, as well as comprehensive cloud identity solutions like Microsoft Entra ID, form the backbone of current identity management for human users and traditional IT systems. Let us discuss their utility and, more importantly, their profound insufficiencies for MAS.

\subsubsection{Lingering Utility for Constrained Scenarios}
In limited contexts, particularly involving single agents or direct human-to-agent platform interactions, these traditional protocols and systems can still play a role, primarily in managing the human interface to agentic systems or bootstrapping initial agent context:
\begin{itemize}
    \item \textbf{Human Authentication to Platforms:} OIDC and SAML for Web/Federated Access: A human user authenticating to an AI agent deployment platform via OIDC or SAML is a standard use case. This is often federated through broader cloud identity solutions like Microsoft Entra ID, which can manage both cloud-native and synchronized enterprise identities. For instance, a developer logging into an AI orchestration platform would use their enterprise OIDC provider. Kerberos for Enterprise Internal Access: Within many corporate networks, Kerberos remains the primary mechanism for authenticating human users to internal services and platforms. A developer or operator might authenticate to their workstation and subsequently to an agent management console using Kerberos.
    \item \textbf{Deriving Initial Agent Context and Attributes:} LDAP as an Attribute Source: Enterprise LDAP directories (such as those underpinning Active Directory, often managed or federated by Microsoft Entra ID in hybrid environments) serve as authoritative sources for user attributes and group memberships. This information can be used by an organization to issue initial Verifiable Credentials (VCs) to an agent, attesting to its ownership, departmental affiliation, or preliminary set of permissions derived from the human deployer's context. The platform may then spawn agents that initially operate under a context derived from this human user's authenticated session. The platform then creates an agent, mcp-dev-agent, which might initially inherit some basic permissions tied to the developer's identity (sourced via OIDC, SAML, or Kerberos, with attributes potentially enriched from LDAP) to access specific code repositories and documentation systems. OAuth 2.0 for Simple Delegated Access by a Single Agent: An AI agent acting as a confidential client can use OAuth 2.0 to access a resource server on behalf of a human user who has granted explicit consent. This mirrors traditional third-party application access. If mcp-dev-agent needs to retrieve additional project context using Model Context Protocol (MCP)~\cite{Anthropic2024MCPNews, MCP2025Spec, narajala2025enterprise} to better understand the developer's codebase, it would go through a standard OAuth 2.0 flow, obtaining an access token scoped specifically to read project documentation and code structures that the developer has authorized.
    \item \textbf{NHI Tasks and Automations:} NHIs can inherit access permissions from the human who deployed them. These identities, while non-autonomous and task-specific, are typically predictable, constrained, and managed through traditional IAM protocols. Service Accounts and OAuth 2.0: Traditional NHIs like service accounts often rely on OAuth 2.0 client credentials flows to authenticate to cloud APIs or internal services. These flows are compatible with existing identity governance platforms, though they lack behavioral awareness and session integrity. Secrets and Certificates as Surrogate Authentication: Static secrets and certificates issued through PKI or secret management systems are effective in authentication but lack real-time behavior verification, traceability, and support in dynamic environments. Role-Based Access Tied to Humans: NHIs in many organizations are indirectly managed by assigning them roles or permissions derived from human owners or creators (e.g., LDAP group inheritance or IAM role mapping). This makes sense for simple automation tools but fails in autonomous systems.
\end{itemize}
However, these scenarios typically involve a single, well-defined agent acting in a relatively static role, often directly tethered to a human user's session or a pre-configured machine identity derived from these traditional IAM systems. The complexities, and the breakdown of these approaches, arise when multiple agents interact autonomously, as detailed next.

\subsubsection{Fundamental Insufficiencies for Multi-Agent Systems (MAS)}
The dynamic, decentralized, and deeply interconnected nature of MAS exposes critical flaws in traditional IAM:
\begin{itemize}
    \item \textbf{Coarse-Grained and Static Permissions:} OAuth and SAML primarily rely on pre-defined scopes or roles that are often too broad and static for the fluid operational needs of AI agents. Agents in MAS frequently require granular, task-specific permissions that can change dynamically based on context, mission objectives, or real-time data analysis~\cite{Koshutanski2008EnhancingGrid}. 
    
    \textit{Example:} Consider a disaster response MAS. Agent-Search (locates survivors via drone\_feed\_api) might initially need read-only access to map data (map.read) and drone telemetry (drone.telemetry.read). Upon finding a survivor, it might need to delegate a task to Agent-MedicalDispatch (coordinates medical\_resources\_api), which then requires access to medical\_assets.request and hospital\_availability.query. Agent-Search might then also need to alert Agent-Logistics (manages supply\_chain\_api) about resource needs, requiring supply.request permissions. In this example, traditional OAuth scopes (read\_all\_data, manage\_all\_resources) would lead to massive over-privileging, while re-authenticating for every micro-permission change is untenable.

    \item \textbf{Single-Entity Focus vs. Complex Delegations:} These protocols are architected around a single authenticated principal (user or application)~\cite{Thompson2007TransitiveAccess}. They struggle to model and secure complex delegation chains where an agent might spawn sub-agents, or where an agent acts on behalf of multiple principals simultaneously (e.g., a user and an organization).

    \textit{Example:} A user (userAlice\_DID) delegates a financial planning task to Agent-Planner (agentPlanner\_DID). Agent-Planner determines it needs specialized market analysis and spawns Agent-MarketAnalyst (agentMarketAnalyst\_DID) and tax optimization from Agent-TaxOptimizer (agentTaxOptimizer\_DID). How is userAlice\_DID's authority securely and granularly passed from Agent-Planner to its sub-agents? Does Agent-MarketAnalyst inherit all of Agent-Planner's (and thus userAlice\_DID's) permissions, or just the bare minimum for market data access? OAuth's delegation (e.g., token exchange) is typically designed for simpler (often one-hop) scenarios and doesn't provide a clear, auditable chain of fine-grained delegated authority. As a result, using the OAuth model, accountability becomes blurred: if Agent-TaxOptimizer accesses unauthorized client data, is Agent-Planner or userAlice\_DID responsible?

    \item \textbf{Limited Context Awareness:} Traditional IAM decisions are largely based on static roles or scopes, with minimal understanding of the runtime context, agent intent, or associated risk level~\cite{Bouhairie2021SCPAC}. Access is often granted at the beginning of a session and persists, irrespective of evolving circumstances.

    \textit{Example:} An inventory management agent (Agent-Inventory) has permissions to update stock levels (inventory.write). If it attempts to update stock levels for a product that has been recalled (an environmental condition) or tries to zero out all inventory (anomalous behavior), traditional IAM systems typically lack the contextual awareness to flag this as suspicious or dynamically restrict the permission.

    \item \textbf{Scalability Issues with Token/Session Management:} For organizations deploying hundreds or thousands of (potentially ephemeral) agents, each potentially interacting with numerous services, the volume of authentication events and tokens can overwhelm traditional IAM infrastructure~\cite{Ren2009NextGenSession}. Managing issuance, validation, and especially revocation of a massive number of short-lived tokens becomes an operational nightmare.

    \textit{Example:} An e-commerce platform deploys thousands of personalized shopping assistant agents for users. In this example, each agent might exist for only a few minutes. The overhead of frequent, secure token management with traditional protocols is a significant barrier.

    \item \textbf{Dynamic Trust Models \& Inter-Agent Authentication:} Agents in MAS often need to authenticate and authorize each other, potentially across organizational boundaries, without a pre-existing, universal trust fabric. OAuth and SAML assume a hierarchical trust model (user trusts IdP, SP trusts IdP). Peer-to-peer trust establishment between autonomous agents from different trust domains is not natively supported~\cite{Shaikh2015TrustIdentity}.

    \textit{Example:} Agent-Alpha from "AlphaCorp" needs to request data processing from Agent-Beta from "BetaInc." How do they mutually authenticate? How does Agent-Beta verify Agent-Alpha's capabilities or authorization to request this specific processing without resorting to cumbersome pre-shared secrets or custom API key mechanisms for every pair of interacting agents?

    \item \textbf{NHI Proliferation and Management Crisis:} Each autonomous agent may require NHIs for numerous APIs, databases, and services, leading to an exponential growth in secrets that must be securely stored, rotated, and managed~\cite{Choda2025SecretSprawl}. This "secret sprawl" increases the attack surface significantly.

    \textit{Example:} A single supply chain optimization agent might need API keys for: a shipping provider, a warehousing system, a customs declaration service, an internal ERP.

    \item \textbf{Global Logout/Revocation Complexity:} If an agent is compromised or its task is complete, ensuring its access rights and sessions are immediately and comprehensively revoked across all systems it interacts with is a major challenge with traditional, often session-based protocols~\cite{Fett2016OAuthAnalysis}. Fragmented revocation mechanisms can leave lingering access.

    \textit{Example:} An agent Agent-DataAggregator has active sessions with three different microservices using OAuth tokens. If the agent is detected as compromised, revoking its token at the authorization server is step one. Ensuring each microservice immediately invalidates its session based on that token, especially if they cache permissions, requires a coordinated effort not always inherent in standard OAuth.
\end{itemize}

\subsection{Unique Challenges Posed by Agentic AI in MAS Further Exacerbating IAM Deficiencies}
Beyond the protocol mismatches, the very nature of agentic AI introduces further complexities:
\begin{itemize}
    \item \textbf{Autonomy and Potential Unpredictability:} Agents with high degrees of autonomy can make decisions that were not explicitly programmed, potentially leading to unforeseen interactions or resource access attempts that challenge static policy definitions.
    \item \textbf{Ephemerality and Dynamic Lifecycles:} Agents can be created, cloned, and destroyed rapidly based on demand. Managing identities and access for such transient entities with persistent credentials is risky and inefficient. An "ephemeral authentication" approach is needed~\cite{SSHCommsJITAuth}.
    \item \textbf{Evolving Capabilities and Intent:} Agents, particularly those incorporating online learning, can adapt their behavior and even their goals over time. An IAM system must be able to accommodate or constrain such evolution.
    \item \textbf{Need for Verifiable Provenance and Accountability:} Tracing actions back to a specific agent instance, understanding its decision-making process (especially if it involved other agents or tools), and ensuring non-repudiation is crucial for trust and forensics.
    \item \textbf{Preventing Autonomous Privilege Escalation:} A sophisticated agent might probe its environment or interact with management APIs to grant itself higher privileges if not carefully constrained. Additionally, agents may interact with each other in a way that leads to privilege escalation through their combined actions, in a manner similar to collusion among humans.
    \item \textbf{Risks of Over-Scoping Access and Permissions:} Agents will actively explore and utilize every permission available to them. This pervasive behavior demands a shift to tightly scoped, task-specific, and context-based access controls to prevent over-privilege and unintended access to sensitive data and environments.
    \item \textbf{Secure and Efficient Cross-Agent Communication \& Collaboration:} As agents increasingly form ad-hoc teams or workflows, the need for secure, low-overhead authentication and authorization between them becomes paramount.
    \item \textbf{Actions Taken May Not Directly Correlate to Human Requests:} As agents are given increasing autonomy and reasoning capabilities, the direct tie between a given human goal and actions taken by any particular agent may no longer exist. For example, a management agent may decide to request a worker agent to use a tool based on its own reasoning, rather than at the specific request of a human. An IAM system must be able to discern between when an action is taken at the direct request of a human, and when it is the result of an agentic decision.
\end{itemize}
These challenges collectively demonstrate that a reactive, bolt-on approach to agent IAM is insufficient. A proactive, purpose-built architectural framework is imperative to harness the power of MAS securely and responsibly. Traditional IAM systems provide a shaky foundation for the towering edifice of interconnected, autonomous AI agents.

\section{Defining the Agent Identity (Agent ID) for a New Era}

To address the challenges of Agentic AI IAM, we must first redefine what constitutes an "identity" for an AI agent. It transcends a simple API key or a username/password. An Agent ID in a MAS context must be a rich, verifiable, dynamic, and cryptographically secured profile that serves as the foundation for trust, access control, and accountability.

\subsection{What Constitutes an AI Agent's Identity? Beyond Static Identifiers}
An AI agent's identity is not merely a label but a comprehensive digital representation that captures its origin, purpose, capabilities, behavior, relationships, and attestations. Agent IDs represent a subset of NHIs that are autonomous, goal-driven, and context-aware. However, to function effectively, agents also rely on or control other types of NHIs (i.e., API tokens, service accounts, workload identities access external resources, execute API calls, or authenticate to services). Agent IDs must be uniquely distinguishable, even when agents are cloned or operate ephemerally, and it must support verification of claims made by or about the agent. We define an "instance" of an AI agent as a runtime instantiation of an agent's software and model, combined with its unique state, memory, and interaction history at a given point in time. Table~\ref{table_agent_identity_models} outlines different identity models for agents based on their lifespan, origin, and hierarchical relationships, highlighting how unique identifiers support traceability and attribution.

\begin{table}[!t]
\renewcommand{\arraystretch}{1.3}
\caption{Agent Identity Models in Multi-Agent Systems}
\label{table_agent_identity_models}
\centering
\begin{tabularx}{\columnwidth}{@{}lX@{}} 
\toprule
\textbf{Agent Type} & \textbf{Description} \\
\midrule
Persistent Agents & For long-lived agents, the ID provides a continuous thread of identity across sessions, state changes, and even restarts, as long as core attributes and memory persist. \\
Ephemeral Agents & Each execution of a short-lived, task-specific agent constitutes a new instance with a unique (potentially derived) ID, ensuring that its actions are distinctly attributable, even if its lifespan is mere seconds. \\
Agent Copies/Forks & A copied or forked agent becomes a distinct instance with its own unique ID, diverging from its parent over time. The relationship to the parent (provenance) should be part of its identity. \\
Hierarchical Agents & Sub-agents spawned by a parent agent are separate instances, each with a unique ID, but with a verifiable link (e.g., via a Verifiable Credential) back to the parent, enabling traceable delegation. \\
\bottomrule
\end{tabularx}
\end{table}

\subsection{Essential Components of an Agent ID}
The proposed Agent ID, ideally anchored by a Decentralized Identifier (DID) \cite{W3C2022DIDs}, should encapsulate a wide array of information within its associated DID Document and through Verifiable Credentials (VCs) \cite{W3C2021VCDataModel1, Sporny2024VCDataModel2}. These components allow for a holistic representation:

\begin{enumerate}[label=(\Alph*)]
    \item \textbf{Cryptographic Anchor \& Verifier:}
    \begin{itemize}
        \item Decentralized Identifier (DID): The globally unique, persistent, and resolvable root identifier (e.g., did:example:agent123). The DID method dictates how it's registered and resolved.
        \item Associated Cryptographic Key Pairs: Public/private key pairs linked to the DID, specified in the verificationMethod section of the DID Document. These are used for signing agent actions, encrypting communications, and authenticating the agent when it presents its DID.
        \item DID Document Service Endpoints: Pointers to services associated with the agent, such as its communication endpoints or a profile service.
    \end{itemize}
    \item \textbf{Core Attributes \& Metadata (Often in DID Document or VCs):}
    \begin{itemize}
        \item Creator/Deployer/Owner/Controller: DIDs or other identifiers of the entities responsible for the agent's creation, operation, and governance.
        \item Agent Software Version \& Model Information: Cryptographic hash of the agent's core model parameters and software version. We recommend the use of FIPS-approved SHA-3 family hash functions (SHA3-224, SHA3-256, SHA3-384, and SHA3-512) to ensure strong cryptographic security.
        \item Timestamps: Creation date, last update, expected expiry (for ephemeral IDs).
        \item Dependencies(Optional): A list of critical software components, libraries, or other agent services that this agent relies upon. This is optional metadata and a normative reference to AIBOM is preferred way to define the dependencies. 
        \item Training Information (Optional): Details about the datasets, methods, and environment used to train the agent's underlying model.
        \item Lifecycle Status: Current state (e.g., active, suspended, revoked, archived).
    \end{itemize}
    \item \textbf{Capabilities, Scope, and Behavior (Crucial for Access Control \& Trust):}
    \begin{itemize}
        \item Formal Scope of Behavior: A machine-readable definition of the agent's intended tasks, operational domains, and interaction boundaries.
        \item Decision-Making Capabilities: Details on the agent's model type, primary reasoning methods, and key behavioral parameters.
        \item Toolset: An explicit, verifiable list of the tools, APIs, or other agents it is authorized to use.
        \item Expected Outcomes \& Limitations: Definition of intended successful outcomes and known failure modes or limitations.
    \end{itemize}
    \item \textbf{Operational \& Security Parameters:}
    \begin{itemize}
        \item Communication Protocols Supported: Specification of protocols the agent can use.
        \item Security Properties Attested: Claims about security features.
        \item Compliance Information: VCs asserting compliance with relevant regulations.
        \item Update Mechanism: Information on how the agent's software, model, or DID Document can be securely updated.
    \end{itemize}
    \item \textbf{Verifiable Credentials (VCs): The Key to Dynamic Attributes and Trust:} VCs are digitally signed attestations about an agent, issued by a trusted entity. Usually, trusted entities are government agencies or big IT companies acting as Certification Authorities. Agents can hold and present these VCs to prove specific attributes or authorizations.
    \begin{itemize}
        \item Role VCs: \url{"DisasterResponseCoordinatorRole".}
        \item Capability VCs:\url{"CertifiedToUse_MedicalImagingAI_v3".}
        \item Reputation VCs: \url{"TrustedCollaborator_Score_95_Percentile_from_CommunityX".}
        \item Provenance VCs: \url{"SpawnedBy_did:example:parentAgent789_at_TimestampZ".}
    \end{itemize}
\end{enumerate}

\subsection{Agent ID Ownership and Control}
A cornerstone of this new IAM paradigm is the principle of Self-Sovereign Identity (SSI) applied to agents.
\begin{itemize}
    \item \textbf{Agent (or its designated controller) as Holder:} The agent itself or its designated controller holds the private keys associated with its DID and manages its VCs.
    \item \textbf{Controller:} The entity ultimately responsible for the agent.
    \item \textbf{Decoupling from Issuers and Verifiers:} The agent’s identity is not solely dependent on a single centralized identity provider.
\end{itemize}
This model moves away from centrally managed identities, empowering the agent/controller with greater control and portability.

\subsection{ID Generation, Assignment, and Lifecycle Management: From Birth to Revocation}
Managing the lifecycle of these rich Agent IDs is crucial.
\begin{itemize}
    \item \textbf{Initial ID Generation and Assignment:}
    \begin{itemize}
        \item Centralized Platform Issuance: In enterprise settings, a platform might generate a DID for an agent upon deployment.
        \item Decentralized/Self-Issuance: An agent or its controller can generate its own DID using a suitable DID method.
        \item Initial Properties: At creation, the DID can be associated with core attributes.
    \end{itemize}
    \item \textbf{Runtime ID Adaptation \& Ephemeral Identities:} Agents may need to operate under different personas or with limited-scope identities for specific tasks.
    \begin{itemize}
        \item Role-Based/Task-Specific IDs: An agent might present a specific VC that grants it a temporary role or use a derived, short-lived DID.
        \item Secure Protocol for Assuming Runtime IDs:
        \begin{enumerate}
            \item Request: The agent requests a new role/ephemeral ID/VC.
            \item Verification: Issuer verifies primary DID and policies.
            \item Issuance: Issuer provides a new (potentially time-bound, scope-limited) VC or ephemeral DID.
            \item Usage: Agent uses the new ID/VC for the specific context.
            \item Revocation/Expiry: The temporary ID/VC is revoked or expires.
        \end{enumerate}
    \end{itemize}
    \item \textbf{ID Update and Revocation:}
    \begin{itemize}
        \item DID Document Updates: Changes to an agent's capabilities or keys require updating its DID Document.
        \item VC Revocation: Invalid VCs must be revoked using mechanisms like VC Status Lists.
        \item DID Deactivation/Revocation: The primary DID can be marked as deactivated if the agent is decommissioned.
    \end{itemize}
\end{itemize}
This rich, dynamic, and verifiable Agent ID serves as the cornerstone of the proposed Agentic AI IAM framework. The demo SDK for Agent ID is published as open source code at Github \cite{Huang2025cAgentDIDSDK}.

\section{The New Agentic AI Identity and Access Management Framework Architecture}

To address the multifaceted challenges of managing AI agents in MAS, we propose a comprehensive IAM framework built upon modern cryptographic primitives and a layered architecture designed for dynamic, secure, and interoperable agent interactions.

\subsection{Foundational Pillars}
The framework rests on several key technological pillars:
\begin{enumerate}[label=(\Alph*)]
    \item \textbf{Decentralized Identifiers (DIDs) and Verifiable Credentials (VCs):} DIDs \cite{W3C2022DIDs} provide globally unique, persistent, cryptographically verifiable identifiers controlled by the agent or its controller, enabling self-sovereign identity essential for cross-organizational and decentralized MAS. VCs \cite{W3C2021VCDataModel1, Sporny2024VCDataModel2} are digitally signed attestations about an agent, allowing granular and dynamic proof of attributes, capabilities, or authorizations. These technologies are particularly well-suited for representing Non-Human Identities (NHIs), which are widely discussed in the industry \cite{OWASP2025NHITop10, CSA2024StateNHISecurity}, providing a standardized approach to managing autonomous agent identities in distributed systems.
    \item \textbf{Zero-Knowledge Proofs (ZKPs):} ZKPs \cite{Goldwasser1989KnowledgeComplexity} allow an agent to prove a statement's truth (e.g., possessing a specific VC attribute) without revealing the underlying information, balancing verifiability with privacy. This is crucial for selective disclosure and proving policy compliance without exposing sensitive internal states.
    \item \textbf{Agent Naming and Discovery Service (ANS):} An ANS, inspired by DNS but tailored for agents, enables secure and reliable discovery based on capabilities, protocols, providers, and versions, not just names \cite{Huang2025ANS}. This could use a naming structure like \url{protocol://AgentFunction.CapabilityDomain.Provider.Version[.protocolExtension]} and resolve to DIDs, with entries secured by PKI or linked to verifiable claims.
\end{enumerate}

\subsection{Core Architectural Layers}

\begin{enumerate}
    \item \textbf{Layer 1: Identity \& Credential Management Layer:} Responsible for creating, issuing, storing, and managing the lifecycle of Agent DIDs and VCs.
    \begin{itemize}
        \item DID Registries/Methods: Systems anchoring DIDs and their DID Documents (e.g., public/permissioned DLTs, did:web, an "Agent ID Provider Network").
        \item VC Issuers and Verifiers: Trusted entities issuing and checking VCs.
        \item Agent Wallets/Secure Storage: Secure agent-side storage for private keys and VCs.
        \item Key Management Services: For key generation, rotation, and revocation.
    \end{itemize}
    \item \textbf{Layer 2: Agent Discovery and Trust Establishment Layer:} Enables agents to find each other and establish trust.
    \begin{itemize}
        \item ANS Resolution Mechanisms: Services implementing the ANS for capability-based discovery.
        \item DID Resolvers: Standard components for retrieving DID Documents.
        \item Reputation Systems: DID-anchored systems for sharing reputation scores.
        \item Trust Frameworks: Policies defining how trust is evaluated (e.g., trusted VC issuers).
    \end{itemize}
    \item \textbf{Layer 3: Dynamic Access Control Layer:} Makes fine-grained, context-aware authorization decisions.
    \begin{itemize}
        \item Policy Decision Point (PDP): Evaluates access requests against policies using agent ID (DID, VCs), resource attributes, action, and context~\cite{OASIS2013XACML}.
        \item Policy Administration Point (PAP): Where policies (e.g., in Rego/OPA~\cite{OPA2025Rego}) are defined~\cite{OASIS2013XACML}.
        \item Policy Information Point (PIP): Gathers attributes for the PDP~\cite{OASIS2013XACML}.
        \item Access Control Mechanisms: ABAC, PBAC, and JIT access using temporary, scoped VCs. 
    \end{itemize}
    \item \textbf{Layer 4: Unified Global Session Management \& Policy Enforcement Layer:} A critical innovation for consistent, real-time establishment, tracking, management, and enforcement of IAM policies, including global logout and session invalidation, across heterogeneous agent communication protocols.
    \begin{itemize}
        \item Cross-Protocol Session Authority (SA): Logically centralized component for global session oversight, policy distribution, orchestrating global logout, and state change propagation.
        \item Adapter Enforcement Middleware (AEM): Lightweight plugins injected into Protocol Adapters, hooking into session initiation, subscribing to SA updates (via SSS), intercepting requests, and enforcing decisions locally, including terminating local sessions on global logout.
        \item Enhanced Protocol Adapters: Gateways understanding specific agent protocols, integrated with AEM for authentication, authorization, and local session management linked to global contexts.
        \item Session State Synchronizer (SSS): Highly available, low-latency distributed data store maintaining a real-time ledger of active global agent session contexts, their mappings to protocol-specific sessions, and current validated capabilities/status. It's the primary source of truth for AEMs regarding session validity.
    \end{itemize}
    \textit{Flow Example: Global Logout for Agent Alpha}
    \begin{enumerate}
        \item Global logout for AgentAlpha\_DID reaches SA.
        \item SA updates SSS: marks GlobalSessionID\_123 (for AgentAlpha\_DID) as "terminated".
        \item SA may push notifications to relevant AEMs.
        \item AEM for A2A adapter, on SSS check (or push), sees termination, invalidates local A2A session.
        \item Similar for MCP adapter's AEM. Further requests from Agent Alpha are blocked.
    \end{enumerate}
\end{enumerate}

\begin{figure}[!t]
\centering
\includegraphics[width=0.65\linewidth]{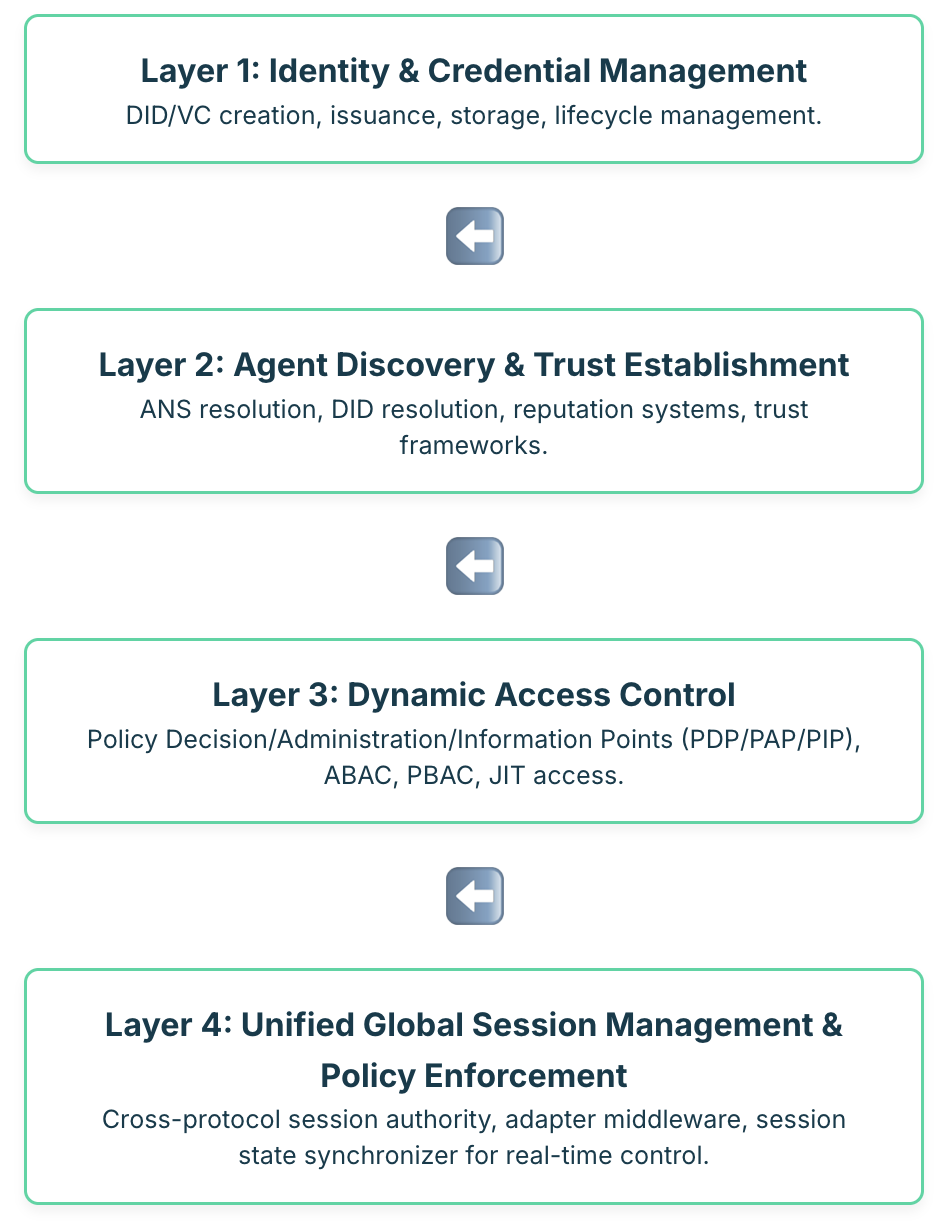} 
\caption{Core Architecture and its layers}
\label{fig:arch}
\end{figure}

\subsection{Applying Zero Trust Principles}
The framework embodies Zero Trust~\cite{Kindervag2010ZeroTrust, Rose2020NISTZTA, llm_genai_security_2025}:
\begin{itemize}
    \item \textbf{Explicit Verification:} Always verify agent identity (DID, VCs) and authorization.
    \item \textbf{Least Privilege Access:} Grant minimum necessary permissions, ideally via JIT VCs.
    \item \textbf{Assume Breach:} Design for compromise; rapid revocation via the Unified Enforcement Layer is key.
    \item \textbf{Micro-segmentation:} Granular agent DIDs support network/application micro-segmentation.
    \item \textbf{Data-Centric Security:} Policies tied to data sensitivity and agent capabilities.
\end{itemize}

\section{Agent IDs in the IAM Process}

This section provides an in-depth exploration of how these constructs enable robust fine-grained access control, ensure secure and non-reputable logging, facilitate effective real-time monitoring and anomaly detection, and empower agile, targeted incident response. A critical enabler for many of these use cases is the Agent Name Service (ANS by \cite{Huang2025ANS}), which provides a secure and capability-aware mechanism for agents to discover each other before interaction. We will illustrate conceptual design patterns, including sample interactions involving emerging agent communication protocols like Google's Agent-to-Agent (A2A) protocol~\cite{Google2025A2AAnnounce} and Anthropic's Model Context Protocol (MCP)~\cite{Anthropic2024MCPNews, MCP2025Spec, Ehtesham2025InteroperabilitySurvey}, demonstrating the framework's adaptability and practical utility in complex Multi-Agent Systems (MAS)\cite{mas_threat_model_2025}. 

\subsection{Fine-Grained Access Control in Action}

Effective access control in MAS must move beyond static roles to embrace dynamic, attribute-based (ABAC), and policy-driven methodologies. The journey often begins with an agent needing to discover another agent or service capable of fulfilling a specific need. This is where the ANS plays a pivotal role, integrated with DIDs and VCs for subsequent secure interaction and authorization.

\textit{Deep Dive into Dynamic Authorization Decisions, Prefaced by ANS Discovery:} Consider TaskOrchestratorAgent (did:com:enterprise:agent:orchestrator:alpha-001) which needs to delegate a financial data analysis task. Its first step is to find a suitable agent. It queries the Agent Name Service (ANS) for an agent that matches certain criteria.

\begin{figure*}[!t]
\centering
\includegraphics[width=0.95\linewidth]{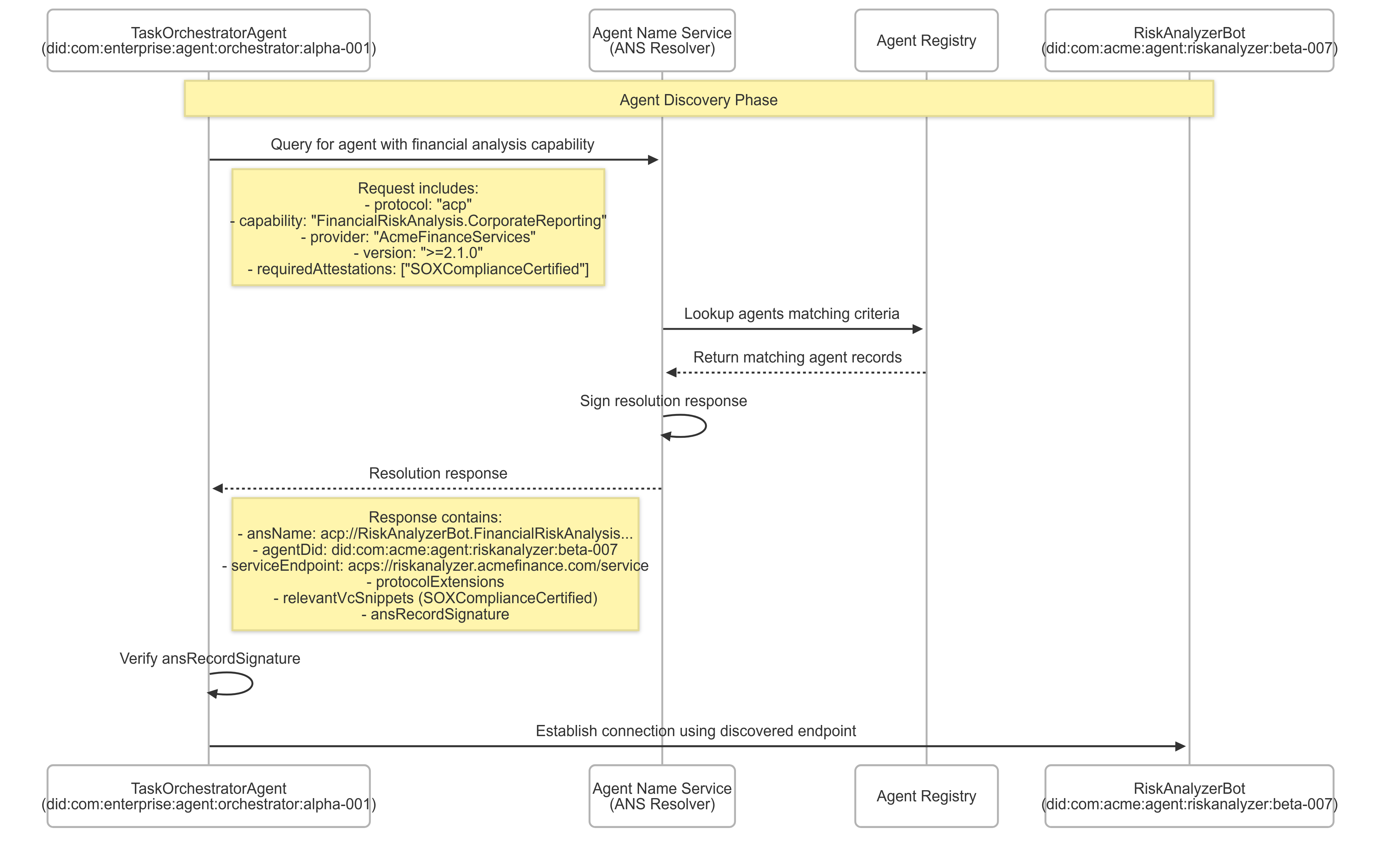} 
\caption{Agent discovery process using the Agent Name Service (ANS)/}
\label{fig:iam1}
\end{figure*}

\subsubsection*{1. ANS Discovery Phase}
TaskOrchestratorAgent constructs an ANS query. The ANS is designed for capability-aware resolution, using a structured naming convention such as: \url{Protocol://AgentID.agentCapability.Provider.vVersion.Extension}.

Conceptual ANS Query (e.g., via a secure API call to an ANS resolver):
\begin{lstlisting}[style=customjson, caption=Conceptual ANS Query, label=lst:ans_query]
// Request to ANS Resolver
{
  "requestType": "resolveAgentByCapability",
  "desiredProtocol": "acp", 
  "requiredCapability": 
    "FinancialRiskAnalysis.CorporateReporting",
  "preferredProvider": "AcmeFinanceServices",
  "versionRange": ">=2.1.0 <3.0.0", 
  "requiredAttestations": [ 
    { "vcType": "SOXComplianceCertified" }
  ]
}
\end{lstlisting}
The ANS resolver (itself a secure, trusted component of the IAM framework, potentially with its own DID and verifiable responses) queries its Agent Registry \cite{Narajala2025ToolSquatting}. The Agent Registry stores information about registered agents, including their ANSNames, DIDs, PKI certificates (if using a PKI-centric ANS as described in your paper), and protocolExtensions detailing their capabilities and associated VCs.

Conceptual ANS Resolution Response:
\begin{lstlisting}[style=customjson, caption=Conceptual ANS Resolution Response, label=lst:ans_response]
// Response from ANS Resolver
{
  "resolutionStatus": "success",
  "resolvedAgents": [
    {
      "ansName": 
        "acp://RiskAnalyzerBot.FinancialRiskAnalysis"
              + ".AcmeFinanceServices.v2.1.3.prod",
      "agentDid": 
        "did:com:acme:agent:riskanalyzer:beta-007",
      "serviceEndpoint": 
        "acps://riskanalyzer.acmefinance.com/service",
      "protocolExtensions": {
        "acp": { "supportedMessagePatterns": 
                  ["request-response", "publish-subscribe"] }
      },
      "relevantVcSnippets": [ 
        { "type": "SOXComplianceCertified", 
          "issuer": "did:com:acme:audit:sox-issuer", 
          "issueDate": "2025-01-15" }
      ],
      "ansRecordSignature": "..." 
    }
    // Potentially other matching agents
  ]
}
\end{lstlisting}
TaskOrchestratorAgent verifies the ansRecordSignature. It now has the DID of a candidate: RiskAnalyzerBot (did:com:acme:agent:riskanalyzer:beta-007).

\begin{figure*}[!t]
\centering
\includegraphics[width=0.95\linewidth]{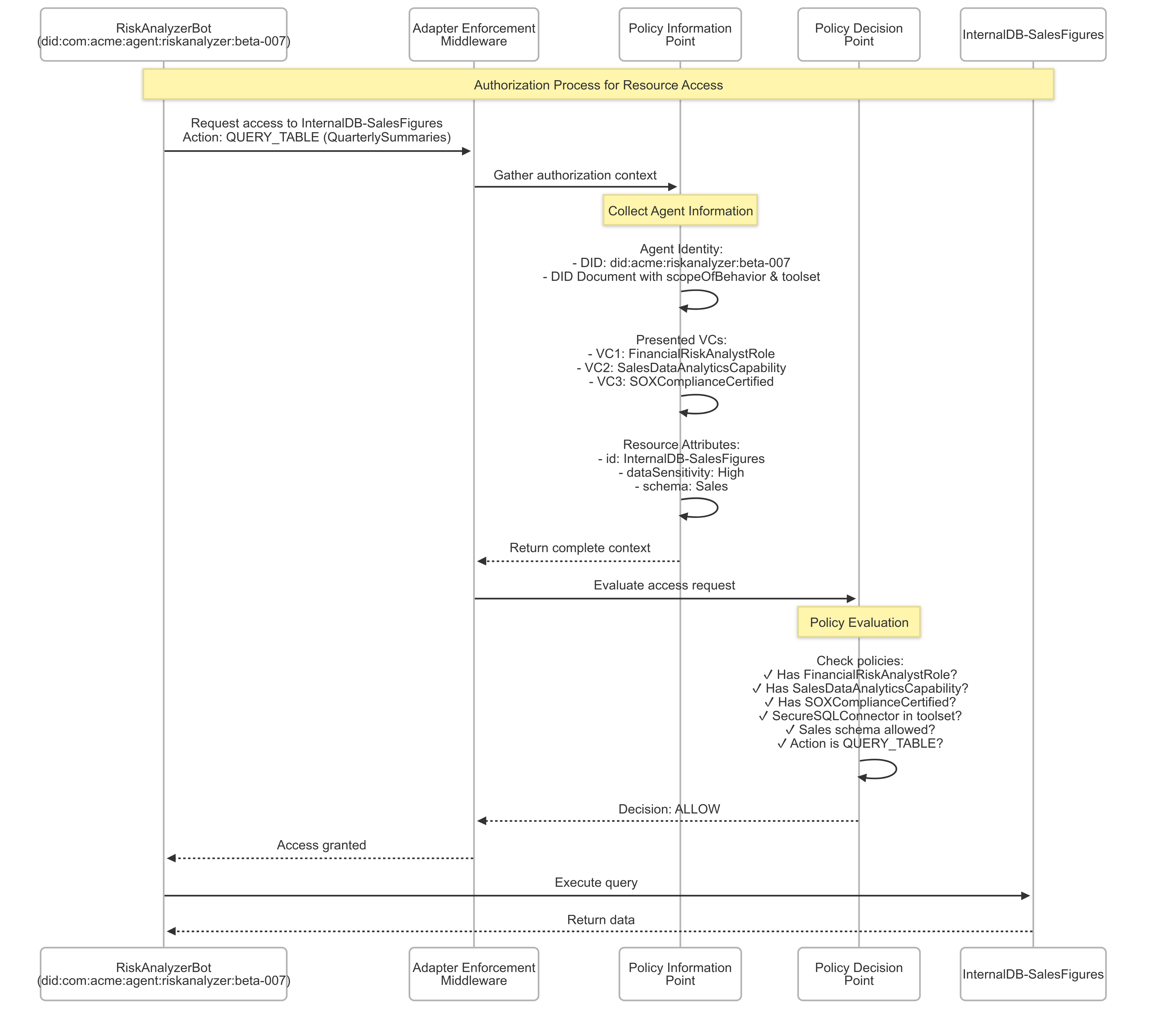} 
\caption{Fine-grained access control enforcement when RiskAnalyzerBot requests access to sensitive financial data.}
\label{fig:iam2}
\end{figure*}

\subsubsection*{2. Interaction and Dynamic Authorization}
TaskOrchestratorAgent now initiates communication with RiskAnalyzerBot (e.g., via ACP). As part of establishing this secure channel or with its first request, RiskAnalyzerBot needs to access InternalDB-SalesFigures and ExternalAPI-MarketSentiment.

The request from RiskAnalyzerBot (let's call it did:acme:riskanalyzer:beta-007) to access InternalDB-SalesFigures is intercepted by the Adapter Enforcement Middleware (AEM, see section IV). The AEM/PIP gathers:
\begin{itemize}
    \item Agent Identity: RiskAnalyzerBot's DID: did:acme:riskanalyzer:beta-007. Its resolved DID Document might state: scopeOfBehavior: "Perform financial risk analysis based on sales and market data." toolset: \{"toolName": "SecureSQLConnector", "targetSchemas": ["Sales", "Projections"]\}.
    \item Presented VCs (obtained during its registration or dynamically):
    \begin{itemize}
        \item VC1 (Role): \{ "type": "FinancialRiskAnalystRole", "issuer": "did:com:acme:hr", ... \}
        \item VC2 (Capability): \{ "type": "SalesDataAnalyticsCapability", "issuer": "did:com:acme:datascience", ... \}
        \item VC3 (SOX Compliance - discovered via ANS): \{ "type": "SOXComplianceCertified", "issuer": "did:com:acme:audit:sox-issuer", ... \}
    \end{itemize}
    \item Resource Attributes: id: InternalDB-SalesFigures, dataSensitivity: High.
    \item Action: QUERY\_TABLE (QuarterlySummaries).
    \item Context: requestTime, sourceIpSegment.
\end{itemize}
The PDP evaluates this against policies. For example:
\begin{lstlisting}[style=customrego, caption=Example Rego Policy for Data Access, label=lst:rego_policy] 
package acme.data_access

default allow = false

# Allow access if agent has correct role, 
# capability VCs, SOX compliance,
# and the requested action is within its 
# declared toolset capabilities for the resource.
allow {
    input.agent.vcs[_].credentialSubject.role == "FinancialRiskAnalystRole"
    input.agent.vcs[_].credentialSubject.capability == "SalesDataAnalyticsCapability"
    input.agent.vcs[_].type[_] == "SOXComplianceCertified" 
    
    # Verify toolset from resolved DID Document 
    # (assuming toolset populated by PIP)
    some tool_idx
    allowed_tool := input.agent.did_document.service[_].
        serviceEndpoint.toolset[tool_idx]
    allowed_tool.toolName == "SecureSQLConnector"
    input.resource.schema IN allowed_tool.targetSchemas # e.g., "Sales"
    input.resource.id == "InternalDB-SalesFigures"
    input.action == "QUERY_TABLE"
    input.resource.table == "QuarterlySummaries" # More granular check
}
\end{lstlisting}
The ANS discovery step ensures that TaskOrchestratorAgent doesn't just find an agent, but finds one that verifiably claims relevant capabilities and compliance (like SOXComplianceCertified) before even attempting interaction. The subsequent authorization then re-verifies these claims (via presented VCs) and checks against more granular policies for resource access. This two-step process (secure discovery then secure, fine-grained authorization) is crucial for building trust and efficiency in large MAS. The DID is the consistent thread linking the discovered entity in ANS to the entity being authorized.

\textit{Just-In-Time (JIT) Access, Enhanced by ANS for Tool Discovery:} Imagine DataProcessingAgent-Temp77 (did:ephemeral:task-xyz:agent-77) is a short-lived agent spawned by WorkflowEngine to perform a specific data transformation. It needs temporary access to a specialized DataTransformationTool-Q.

ANS for Tool Discovery: WorkflowEngine (or DataProcessingAgent-Temp77 itself if it has this capability) first queries the ANS to discover a suitable and currently available instance of DataTransformationTool-Q. ANS Query:
\begin{lstlisting}[style=customjson, caption=ANS Query for Tool Discovery, label=lst:ans_tool_query]
{
  "requestType": "resolveAgentByNameAndCapability", 
  "ansNamePattern": "mcp://DataTransformationTool-Q.*"
        + ".AcmeTools.v1.*.internal", 
  "requiredCapability": "VectorEmbeddings.HighDimReduction",
  "availabilityRequirement": "online_accepting_jobs" 
}
\end{lstlisting}
The ANS returns the DID of an available instance, e.g., did:com:acmetools:mcp:tool:transformQ:instance03.

JIT VC Issuance via MCP Context (Conceptual): WorkflowEngine (acting as a trusted issuer for this context) issues a JIT VC to DataProcessingAgent-Temp77:
\begin{lstlisting}[style=customjson, caption=JIT Verifiable Credential for MCP Tool Access, label=lst:jit_vc_mcp]
{
  "type": ["VerifiableCredential", "MCPToolAccessPass"],
  "issuer": "did:com:acme:workflow:engine-issuer",
  "validFrom": "2025-10-02T14:30:00Z",
  "validUntil": "2025-10-02T14:45:00Z", // Valid for 15 mins
  "credentialSubject": {
    "id": "did:ephemeral:task-xyz:agent-77",
    "authorizedToolDID": "did:com:acmetools:mcp:"
        + "tool:transformQ:instance03",
    "allowedActions": ["executeTransform"],
    "inputDataHandle": "blob://temp-input-xyz",
    "outputDataHandle": "blob://temp-output-xyz",
    "jobId": "job-ephemeral-77a"
  }
}
\end{lstlisting}

\begin{figure*}[!t]
\centering
\includegraphics[width=0.95\linewidth]{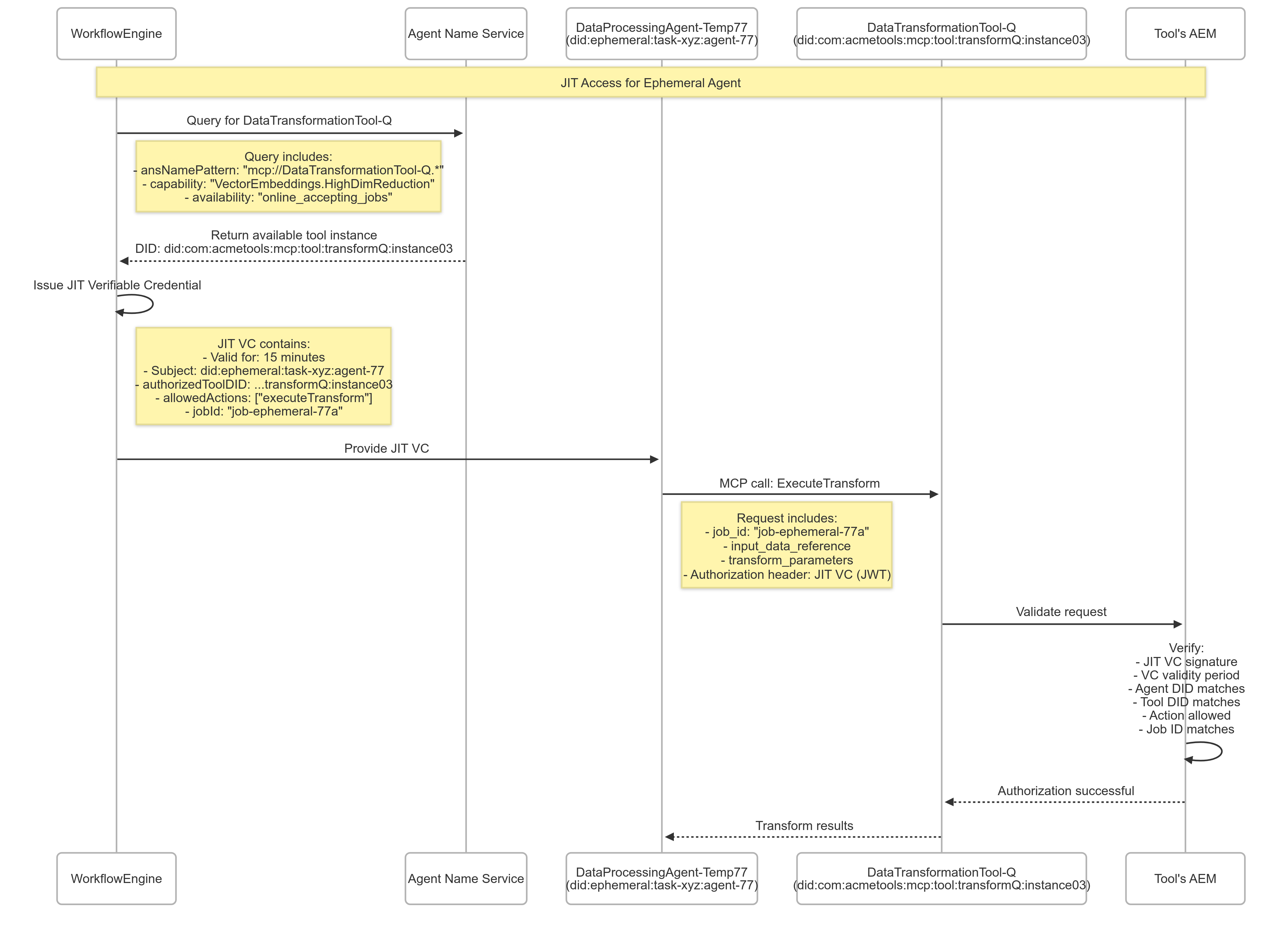} 
\caption{Ephemeral agent authorization using Just-In-Time (JIT) Verifiable Credentials for Model Context Protocol (MCP) tool access.}
\label{fig:iam3}
\end{figure*}

MCP Tool Invocation with JIT VC: DataProcessingAgent-Temp77 invokes DataTransformationTool-Q (whose MCP endpoint was found via ANS then DID resolution). It presents this JIT VC within the MCP call.

Conceptual MCP Call (e.g., using gRPC or HTTP, carrying VC in metadata/headers): Let's assume MCP uses gRPC and metadata for auth as customized transport.
\begin{lstlisting}[style=customcode, caption=Conceptual MCP Tool Call with JIT VC, label=lst:mcp_tool_call]
// Conceptual .proto definition for an MCP tool call
service TransformationTool {
  rpc ExecuteTransform(TransformRequest) returns (TransformResponse);
}

message TransformRequest {
  string job_id = 1;
  string input_data_reference = 2; // "blob://temp-input-xyz"
  map<string, string> transform_parameters = 3;
}

// Client-side pseudocode for DataProcessingAgent-Temp77:
# Assume 'mcp_tool_stub' is the gRPC stub for DataTransformationTool-Q
# Assume 'jit_vc_jwt' is the JIT VC serialized as a JWT
metadata = [
    ('x-agent-did', 'did:ephemeral:task-xyz:agent-77'),
    ('authorization-vc', jit_vc_jwt) 
] # gRPC metadata
request_payload = TransformRequest(
    job_id="job-ephemeral-77a",
    input_data_reference="blob://temp-input-xyz",
    transform_parameters={"algorithm": "PCA", "dimensions": 128}
)
try:
    response = mcp_tool_stub.ExecuteTransform
    (request_payload, metadata=metadata)
    # Process response and write to "blob://temp-output-xyz"
except grpc.RpcError as e:
    # Handle authorization failure or tool error
    log(f"MCP tool call failed: {e.details()}")
\end{lstlisting}
Verification at MCP Tool's AEM: The AEM for DataTransformationTool-Q extracts and verifies the DID and jit\_vc\_jwt. The PDP checks if did:ephemeral:task-xyz:agent-77 is authorized by this specific VC to call this tool instance (did:com:acmetools:mcp:tool:transformQ:instance03) for executeTransform with the given jobId and data handles, and if the VC is within its validity period.

ANS helps find the right instance of a potentially multi-instance MCP tool. The JIT VC then provides extremely narrow, time-bound permission for that specific job and data, dramatically reducing risk compared to the ephemeral agent having broader, longer-lived credentials for a generic tool type.

\textit{Capability-Driven Authorization with A2A Protocol:} AlertingAgent-SystemX (did:com:sysX:a2a:alerter:main:v1) needs to send a critical security alert to a SOCDashboardAgent-PlatformY (did:com:platY:a2a:socdash:primary:v2).

ANS Discovery: AlertingAgent-SystemX resolves \url{a2a://SOCDashboardAgent.SecurityAlertIngestion.PlatformY.v2.critical} via ANS to find the DID and A2A endpoint of SOCDashboardAgent-PlatformY. The ANS response might also indicate that the SOC agent requires alerts to be signed with a key whose DID is on an approved list.

A2A Message Construction with IAM Context: AlertingAgent-SystemX holds a VC: \{"type": "CriticalAlertSourceCredential", "issuer": "did:com:sysX:security-authority", "credentialSubject": \{"id": "did:com:sysX:a2a:alerter:main:v1", "authorizedAlertTypes": ["SECURITY\_CRITICAL", "SYSTEM\_DOWN"]\}\}.

Conceptual A2A Message from AlertingAgent-SystemX (JSON-like payload for an A2A message):
\begin{lstlisting}[style=customjson, caption=Conceptual A2A Message with IAM Context, label=lst:a2a_message]
{
  "a2aHeader": { 
    "messageId": "msg-uuid-9876",
    "senderId": "did:com:sysX:a2a:alerter:main:v1", 
    "recipientId": "did:com:platY:a2a:socdash:primary:v2",
    "protocolVersion": "A2A/1.0"
  },
  "iamExtension": { 
    "verifiablePresentation": [ /* JWT of CriticalAlertSourceCredential */ ],
    "messageSignature": { 
      "keyId": "did:com:sysX:a2a:alerter:main:v1#key-1", 
      "algorithm": "EdDSA",
      "signatureValue": "..." 
    }
  },
  "payload": {
    "alertType": "SECURITY_CRITICAL",
    "sourceSystem": "SystemX_Firewall_Cluster",
    "details": "Multiple intrusion attempts detected from IP range Z.Z.Z.Z",
    "severity": 5, 
    "timestamp": "2025-10-02T15:00:10Z"
  }
}
\end{lstlisting}
Many emerging A2A protocols are defining ways to carry security contexts, often leveraging JWTs or similar token formats within their headers or as part of the message envelope. The iamExtension is a way our framework's specific needs (DID, VP) can be mapped.

\begin{figure*}[!t]
\centering
\includegraphics[width=0.95\linewidth]{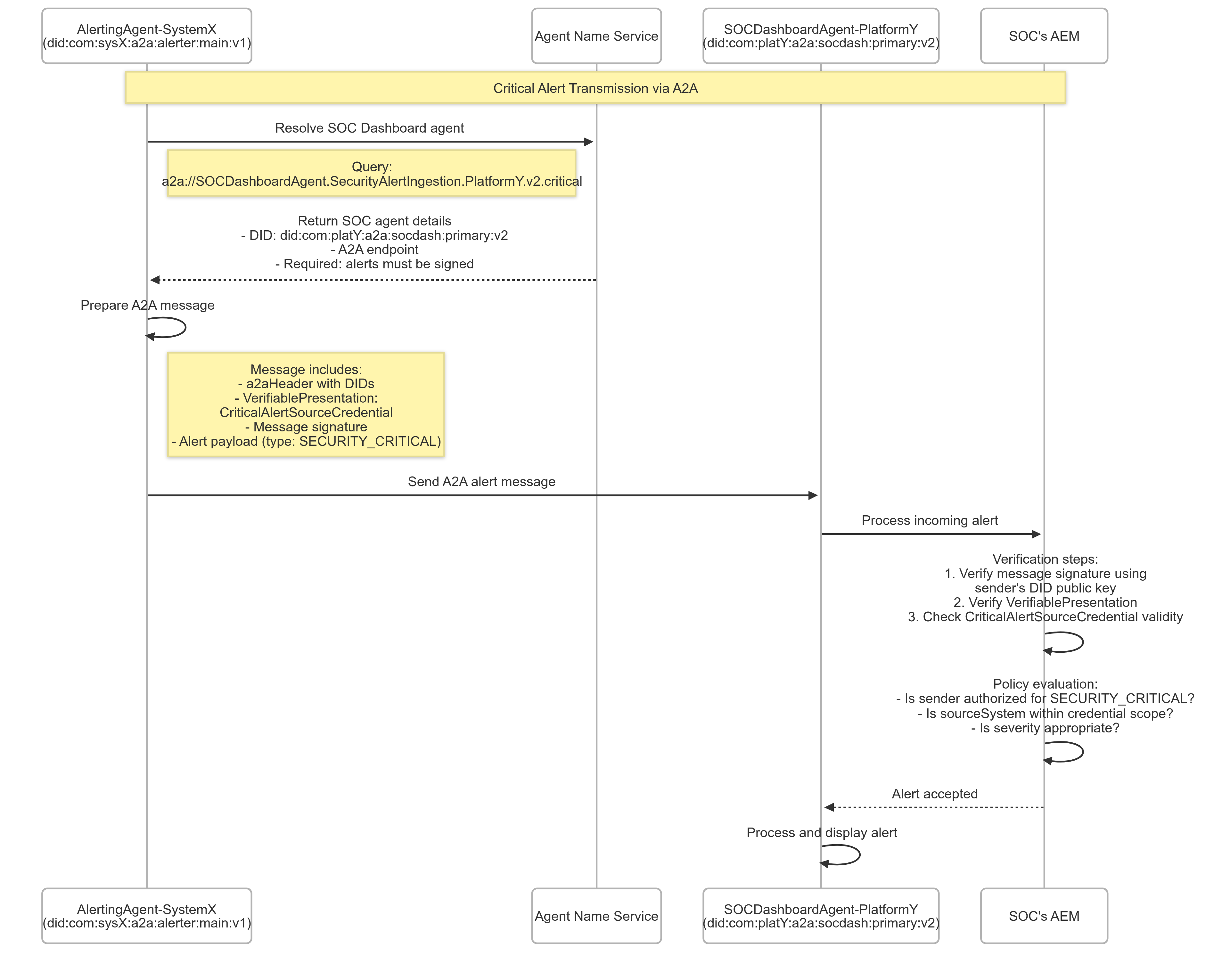} 
\caption{Secure agent-to-agent communication using Google's A2A protocol for critical security alerts.}
\label{fig:iam4}
\end{figure*}

Processing at SOCDashboardAgent-PlatformY's AEM:
\begin{itemize}
    \item AEM verifies messageSignature using the public key from did:com:sysX:a2a:alerter:main:v1\#key-1 (resolved via DID document).
    \item AEM verifies the verifiablePresentation containing the CriticalAlertSourceCredential.
    \item PDP checks policies like: "Accept SECURITY\_CRITICAL alert IF sender DID holds valid CriticalAlertSourceCredential AND the alert's declared sourceSystem is within the scope covered by that credential."
\end{itemize}
The ANS ensures AlertingAgent-SystemX reliably finds the authentic SOCDashboardAgent-PlatformY (not an imposter). The VC presented proves the sender is authorized to issue critical alerts, and the message signature ensures integrity and non-repudiation for the alert content. This provides much stronger guarantees than simple IP whitelisting or pre-shared API keys between agents for A2A communication\cite{securing_a2a}.

The use of ANS for initial discovery, followed by DID-based authentication and VC-based authorization at the point of interaction, forms a robust sequence for secure and fine-grained access control in diverse MAS scenarios.

The Identity and Access Management (IAM) framework for multi-agent systems operates through a sophisticated four-phase lifecycle that ensures secure agent discovery, authentication, and authorization across diverse protocol environments. As illustrated in Figure~\ref{fig:iam1}, the process begins with capability-aware agent discovery through the Agent Name Service (ANS), where requesting agents query for specific capabilities, compliance requirements, and protocol preferences, receiving cryptographically signed responses containing target agent DIDs, service endpoints, and relevant attestations such as SOX compliance certifications. Once agents establish contact, the framework employs dynamic, attribute-based access control as demonstrated in Figure~\ref{fig:iam2}, where the Adapter Enforcement Middleware (AEM) coordinates with Policy Information Points (PIP) to gather comprehensive agent context including Decentralized Identifiers (DIDs), Verifiable Credentials (VCs), and resource attributes, before Policy Decision Points (PDP) evaluate fine-grained authorization policies that consider roles, capabilities, toolset permissions, and data sensitivity levels. For ephemeral or temporary agents, the framework supports Just-In-Time (JIT) credential issuance as shown in Figure~\ref{fig:iam3}, where workflow engines discover available tools via ANS and issue time-limited Verifiable Credentials with narrow, job-specific permissions that enable secure Model Context Protocol (MCP) interactions while minimizing credential exposure and attack surfaces. Finally, the framework facilitates secure inter-agent communication through emerging protocols like Google's Agent-to-Agent (A2A) protocol, as depicted in Figure~\ref{fig:iam4}, where agents exchange cryptographically signed messages containing verifiable presentations that prove authorization for specific communication types, ensuring non-repudiation and enabling real-time security alerting between heterogeneous agent platforms while maintaining end-to-end trust and accountability throughout the multi-agent ecosystem.

\subsection{Secure Logging, Auditing, and Non-Repudiation}

In systems where autonomous agents perform significant actions, establishing a clear, trustworthy, and irrefutable record of events is paramount. This section delves into how the proposed IAM framework, leveraging rich Agent IDs (DIDs and VCs) and the Agent Name Service (ANS) for discoverable context, transforms logging into a critical component of system integrity, accountability, and auditability.

\textit{Immutable Agent Identifiers (DIDs) as the Linchpin of Audit Logs:} Every significant action initiated or participated in by an agent MUST be logged with its unique, persistent Decentralized Identifier (DID) as the primary subject identifier. This creates an unambiguous, globally unique, and cryptographically verifiable link to the specific agent instance responsible for any given event.

\textit{Enhanced Log Granularity with DID and VC Context:} Beyond simply logging the agent's DID, comprehensive logs should capture:
\begin{itemize}
    \item Precise Timestamp: Synchronized across the MAS to ensure correct event sequencing.
    \item Agent DID and ANSName: Logging both the DID (for cryptographic verifiability) and the resolved ANSName (e.g., \url{acp://RiskAnalyzerBot.FinancialRiskAnalysis.AcmeFinanceServices.v2.1.3.prod}) provides human-readable context about the agent's role and origin.
    \item Target Resource(s) DIDs/ANSNames: If the interaction target is another agent or a resource registered in ANS, its DID and ANSName should also be logged.
    \item Input Parameters/Data Hashes: Hashing critical inputs helps reconstruct the context of an agent's decision without necessarily storing sensitive raw data in logs.
    \item Specific Verifiable Credentials (VCs) Presented: The unique identifiers (e.g., id or transaction\_id) of all VCs presented by the agent to authorize that specific action. For example, logging vc:jwt:uri:issuer-finance-bob:task-q3report2025-instance-002 allows an auditor to later retrieve and verify this exact VC.
    \item DIDs and ANSNames of Collaborating Agents: In multi-agent tasks, the DIDs/ANSNames of all significant contributing agents should be logged to trace collaborative decision-making.
    \item Outcome and Policy Reference: The result of the action and a reference to the specific policy version (e.g., ACME\_Finance\_Policy\_v3.2.1\_Rule7) that permitted it.
\end{itemize}
Example Enriched Log Entry incorporating ANSNames:
\begin{lstlisting}[style=customjson, caption=Enriched Log Entry with ANSNames, label=lst:enriched_log]
{
  "eventId": "evt_20251002T110530Z_A789F123",
  "timestamp": "2025-10-02T11:05:30.123Z",
  "initiatingSystem": "WorkflowOrchestratorInternal",
  "agentDid": "did:com:acme:agent:riskanalyzer:beta-007",
  "agentAnsName": 
    "acp://RiskAnalyzerBot.FinancialRiskAnalysis"
    + ".AcmeFinanceServices.v2.1.3.prod",
  "actionPerformed": "ExecuteSecureSQLQuery",
  "targetResourceDid": 
    "did:com:acme:resource:db:InternalDB-SalesFigures",
  "targetResourceAnsName": 
    "db://InternalDBSales.FinancialData"
        + ".AcmeInternal.v1.prod", 
  "inputParametersHash": "sha256-c4d5e6f...",
  "presentedVcIds": [
    "vc:jwt:uri:acme-hr:role-finanalystL2-inst-001", 
    "vc:jwt:uri:acme-audit:sox-compliance-inst-003"
  ],
  "decisionPolicyId": 
    "ACME_DataAccess_Policy_v1.7_Rule12b",
  "collaborationContext": {
    "triggeringAgentDid": 
    "did:com:enterprise:agent:orchestrator:alpha-001",
    "triggeringAgentAnsName": "acp://TaskOrchestrator.CoreBusinessLogic"
        + ".AcmeEnterprise.v1.0.main",
    "taskId": "task_QuarterlyRiskAssessment_2025Q3"
  },
  "outcome": { "status": "Success", "rowsAffected": 0, "dataRetrievedHash": "sha256-g7h8i9j..." },
  "logEntrySignature": "..." 
}
\end{lstlisting}
Logging ANSNames alongside DIDs makes logs instantly more interpretable for human auditors. The cryptographic link via DIDs ensures the identifier is not just a mutable string. The logged VCs provide the exact authorization context for the action, making audits far more precise.

\textit{Cryptographic Non-Repudiation of Agent Actions via DID Signatures:} To achieve strong non-repudiation, critical agent actions or the data they produce can be digitally signed by the agent using the private key associated with its DID. This is particularly important for actions with financial, legal, or safety implications.

Scenario (A2A Context): OrderPlacementAgent (did:com:retail:a2a:orderbot:v1.0, ANSName \url{a2a://OrderPlacement.RetailTransactions.MegaCorp.v1.0.live}) submits a purchase order to SupplierFulfilmentAgent (did:com:supplierX:a2a:fulfill:v2.1, ANSName \url{a2a://Fulfilment.SupplyChain.SupplierX.v2.1.prod}), which was discovered via ANS query for "SupplyChain.OrderFulfilment.SupplierX".

A2A Message with Signed Payload and DID Context: The OrderPlacementAgent constructs an A2A message. The core business payload (the order details) is signed.
\begin{lstlisting}[style=customjson, caption=A2A Message with Signed Payload, label=lst:a2a_signed_payload]
// A2A Message (Conceptual JSON representation)
{
  "a2aHeader": {
    "messageId": "order-uuid-554433",
    "senderId": "did:com:retail:a2a:orderbot:v1.0", 
    "recipientId": "did:com:supplierX:a2a:fulfill:v2.1", 
    "protocolVersion": "A2A/1.0",
    "timestamp": "2025-10-02T16:30:00Z"
  },
  "iamExtension": { 
    "verifiablePresentation": [ /* Optional: JWT of a relevant VC */ ]
  },
  "payload": { 
    "orderId": "PO-2025-10-778",
    "items": [ {"sku": "XYZ123", "quantity": 100}, {"sku": "ABC789", "quantity": 50} ],
    "shippingAddress": "123 Main St, Anytown",
    "totalAmount": 12500.75,
    "currency": "USD"
  },
  "payloadSignature": { 
    "keyId": 
        "did:com:retail:a2a:"
        + "orderbot:v1.0#key-transact", 
    "algorithm": "EdDSA",
    "signatureValue": "..." 
  }
}
\end{lstlisting}
Verification and Logging by SupplierFulfilmentAgent:
\begin{itemize}
    \item The AEM at SupplierFulfilmentAgent's side first authenticates the sender via its DID and any presented VCs (as per Section V-A).
    \item It then specifically verifies the payloadSignature using the public key did:com:retail:a2a:orderbot:v1.0\#key-transact (obtained by resolving the sender's DID).
    \item SupplierFulfilmentAgent's log entry for receiving this order would include: its own DID/ANSName, the sender's DID/ANSName, the order ID, a hash of the received payload, and the payloadSignature object. This creates a verifiable record that OrderPlacementAgent indeed sent that specific order. The initial discovery via ANS ensures the order is sent to a legitimate fulfilment agent. The DID-based signature on the payload provides strong non-repudiation for the order's content, traceable to a specific, verifiable agent identity. Traditional EDI or API calls often rely on weaker authentication or channel security alone.
\end{itemize}

\textit{Verifiable Provenance Chains in MCP Tool Interactions:} When an LLM-based agent uses an MCP tool, understanding the full chain—from user prompt to LLM, to MCP tool call, to tool result, back to LLM, and then to the user—is vital for auditing and debugging.

Scenario: A user asks ResearchLLM-Agent (did:com:ai-lab:mcp:researcher:zeta:v3.1, ANSName \url{mcp://Researcher.ScientificQuery.AILab.v3.1.experimental}) a complex question requiring a database lookup via an MCP tool, SemanticSearchTool (did:com:datastore:mcp:tool:semsearch:v1.0, ANSName \url{mcp://SemanticSearch.KnowledgeBase.DataCorp.v1.0.main}). ResearchLLM-Agent discovers SemanticSearchTool via an ANS query specifying the "KnowledgeBase.SemanticSearch" capability.

MCP Interaction Logging with DIDs and VCs:
\begin{itemize}
    \item User Interaction Log: User prompt, timestamp, and ResearchLLM-Agent's DID/ANSName.
    \item ResearchLLM-Agent Internal Log (or trace):
    \begin{itemize}
        \item Decision to use SemanticSearchTool.
        \item Query sent to ANS for SemanticSearchTool.
        \item Resolved DID/ANSName for SemanticSearchTool.
        \item The MCP call it constructs to SemanticSearchTool, including:
        \begin{itemize}
            \item Its own DID as the caller.
            \item The JIT VC it obtained/presented for this tool use (e.g., vc:jwt:...:mcp-tool-access-zeta-job778).
            \item The parameters sent to the tool.
        \end{itemize}
        \item This entire MCP call could be signed by ResearchLLM-Agent.
    \end{itemize}
    \item SemanticSearchTool (MCP Tool) Log:
    \begin{itemize}
        \item Its own DID/ANSName.
        \item Receiving the MCP call from ResearchLLM-Agent (DID/ANSName logged).
        \item The presented JIT VC ID.
        \item Verification status of the caller's DID and VC.
        \item Parameters received.
        \item Actions it took (e.g., database queries it made internally).
        \item The result it returned to ResearchLLM-Agent.
        \item This log entry or the response payload could be signed by SemanticSearchTool.
    \end{itemize}
    \item ResearchLLM-Agent Internal Log (continued):
    \begin{itemize}
        \item Response received from SemanticSearchTool (potentially with signature verification).
        \item How it processed the tool's output.
        \item The final answer generated for the user (this answer could also be signed).
    \end{itemize}
\end{itemize}
This chained logging, where each step is linked by verifiable DIDs/ANSNames and specific VCs or signed messages, creates a rich, end-to-end auditable provenance trail. If the final answer is wrong, auditors can trace back: Was it the LLM's reasoning, the MCP tool's execution, the data the tool accessed, or the initial ANS discovery that pointed to an incorrect tool version? This detailed, verifiable chain is crucial for explainability and accountability in complex agentic workflows involving external tools.

\textit{Privacy-Preserving Audits of IAM Policies with ZKPs:} Organizations may need to prove to external auditors or regulators that their Agentic AI IAM policies are being correctly enforced, without revealing the proprietary details of all policies or all agent interactions.

Scenario: An auditor wants to verify that access to resources tagged PII\_Strict is only ever granted if an agent presents a valid VC of type PII\_AccessLevel3\_Certified and the request originates from an approved network segment.
Mechanism:
\begin{itemize}
    \item The IAM system's Policy Decision Point (PDP) logs all its decisions, including the agent DID, resource, action, presented VCs (or their hashes), contextual attributes, and the allow/deny outcome. These logs themselves could be cryptographically committed to (e.g., a hash chain).
    \item The organization can run a process that analyzes these logs and generates a ZKP. This ZKP would prove a statement like: "For all access requests to resources tagged PII\_Strict within the last audit period that resulted in an 'allow' decision, the requesting agent's presented credentials included a valid (non-revoked, correctly signed) PII\_AccessLevel3\_Certified VC from an approved issuer, AND the source network attribute was in the set \{'segA', 'segB'\}."
    \item This ZKP is generated without revealing the specific agent DIDs, resource DIDs, exact times, or other details of the individual access events.
    \item The auditor receives and verifies this ZKP, along with information about the approved VC issuers and network segments, providing strong assurance of policy enforcement without seeing the raw, potentially sensitive log data. This enables "compliance as code" verification with privacy. It allows organizations to demonstrate adherence to internal or external IAM rules without exposing the minutiae of every transaction, which is a common challenge in traditional audit processes that often require extensive (and risky) data sharing.
\end{itemize}
By deeply integrating verifiable Agent IDs (DIDs/VCs), secure discovery via ANS, and cryptographic techniques like digital signatures and ZKPs into the logging and auditing process, our framework aims to create a system where agent actions are not just recorded, but are verifiably attributable, contextualized, and, where necessary, proven compliant in a privacy-respecting manner. This robust auditability is fundamental to building and maintaining trust in complex and autonomous MAS.

\subsection{Real-time Monitoring and Anomaly Detection}

Effective IAM extends beyond static policy enforcement to encompass continuous, real-time oversight of agent activities. The rich, verifiable Agent IDs (DIDs and VCs), coupled with the contextual information available through Agent Name Service (ANS) resolutions, provide the foundation for a far more sophisticated and proactive monitoring and anomaly detection capability than achievable with traditional, opaque identifiers. This allows security systems to not only identify what is happening but also understand if it aligns with an agent's intended and attested purpose and capabilities.

\textit{Establishing Rich Behavioral Baselines Anchored to Verifiable Identities (DIDs and ANSNames):} Modern monitoring can move beyond tracking simple metrics like CPU usage per IP address. The proposed framework allows for the creation of multifaceted behavioral baselines for each unique agent DID and its associated ANSName profiles:
\begin{itemize}
    \item \textbf{Discovered vs. Declared Scope of Behavior:} The agent’s DID Document contains its scopeOfBehavior (e.g., "customer\_support\_query\_resolution\_for\_product\_X"). ANS registration might also include a primary capability (e.g., Support.ProductQuery.CustomerFacing.v1). Monitoring systems can compare the agent's actual interactions and data access patterns against this declared and discoverable purpose. Significant deviations trigger alerts.
    
    \textit{Scenario:} SupportAgentAlpha (did:com:support:agent:alpha01, ANSName \url{helpdesk://Support.ProductQuery.CustomerFacing.v1.Acme}) normally accesses the product knowledge base and customer ticket system. If it suddenly starts making frequent ANS queries for agents with FinancialData.InternalAudit capabilities, or attempts to access database schemas related to payroll, this is a strong anomaly relative to its declared/discovered scope.
    
    \item \textbf{Authorized Toolset and ANS-Discoverable Service Usage:} The agent's DID Document details its toolset (specific APIs, other agent DIDs/ANSNames it's authorized to interact with). Monitoring systems can track: Actual tool/API calls made. ANS queries made by the agent to discover other services. If the agent attempts to use tools not in its list or interact with DIDs/ANSNames that don't match its typical collaboration patterns or authorized interaction VCs.
    
    \textit{Scenario (MCP Context):} DataPipelineAgent-ETL (did:com:dataops:agent:etl04, ANSName \url{mcp://ETL.DataWarehouseLoading.DataOps.v2.nightly}) is authorized to use PostgresConnectorTool (an MCP tool discovered via ANS as \url{mcp://DBConnector.PostgreSQL.InternalTools.v1.stable}) and S3StorageTool. If it makes an ANS query for \url{mcp://ExternalAPI.SocialMediaScraping...} or attempts to invoke such a tool via MCP, it's a policy violation and an anomaly.
    
    \item \textbf{VC Presentation Patterns:} Monitoring the types of VCs an agent typically presents for different actions, and the issuers of those VCs. An agent suddenly presenting a VC from a previously unseen or untrusted issuer for a high-privilege operation is suspicious.
    
    \item \textbf{Communication Graph and Trust Dynamics:} Building a graph of typical agent-to-agent interactions (DID-to-DID or ANSName-to-ANSName) based on historical communication logs. New, unexpected communication links, especially with agents outside the organization or with low reputation scores (if a reputation system is integrated), can be flagged.
    
    \textit{Scenario:} A fleet of InventoryCheckAgent instances (e.g., \url{a2a://InventoryCheck.RetailStoreXYZ.Ops.v1.hourly::did:...}) typically only communicate via A2A with a central InventoryMasterAgent (\url{a2a://InventoryMaster.HeadOffice.Ops.v3.main::did:...}). If one InventoryCheckAgent initiates an A2A connection to an unknown external ANSName/DID, or starts sending unusually large A2A payloads, this is anomalous.
\end{itemize}

\textit{Advanced Deviation Detection Leveraging Verifiable Claims:} The ability to verify claims presented as VCs in real-time enhances anomaly detection:
\begin{itemize}
    \item \textbf{Scope Creep Beyond VC-Attested Capabilities:} An agent, ResearchSummarizer (did:..., ANSName \url{a2a://Summarization.ScientificLiterature.ResearchGroup.v1}), might hold a VC for "Access\_PubMed\_API\_SummarizationOnly." If it attempts to use the PubMed API's "BulkDownloadAbstracts" function (which its VC does not authorize), the AEM/PDP would block it, and the monitoring system would log this as a significant deviation, as it's attempting an action beyond its attested capability.
    \item \textbf{Anomalous JIT VC Requests:} If an agent frequently requests JIT VCs for tasks outside its typical operational parameters, or if the requested scopes for JIT VCs escalate without justification, this could indicate a compromised agent or a misbehaving workflow.
    \item \textbf{Interaction with Agents Lacking Expected Counter-Attestations:} If SecureDataTransferAgent is only supposed to send data to other agents that can present a "DataRecipient\_EncryptionLevel5\_Compliant" VC, an attempt to send data to an agent (discovered via ANS) that cannot present such a VC would be a flagged anomaly, even if basic network connectivity is possible.
\end{itemize}

\textit{Dynamic Trust Scoring and Risk-Adaptive IAM Incorporating ANS Context:} The Agent ID (DID) becomes the anchor for a dynamic trust score, influenced by monitoring. ANS context adds another layer.
\begin{itemize}
    \item \textbf{Inputs to Trust Score (with ANS context):} Successful completion of tasks within the agent's ANS-declared capability. Policy violations or anomalous behaviors (as detailed above). Validity and issuer trustworthiness of its VCs. Feedback from other reputable agent DIDs (whose own ANS profiles might indicate their roles/trustworthiness). ANS-related anomalies: Repeatedly querying ANS for unrelated capabilities, attempting to register with a misleading ANSName, or interacting with agents resolved from suspicious ANS domains.
    \item \textbf{Risk-Adaptive Policy Enforcement Example (A2A):} PaymentAgent-Acquirer (\url{a2a://PaymentProcessing.Acquisition.FinServ.v2.live::did:...}) normally processes transactions. It starts making unusual ANS queries for \url{a2a://DataAggregation.UserProfiling...} services and receives a few low-severity alerts for attempting to access non-payment related internal APIs. Its trust score, managed by the IAM system, is lowered. The Session Authority (SA) is notified of the trust score change. The SA updates the Session State Synchronizer (SSS) for this agent's global session, adding a "ReducedTrust" status or dynamically adjusting its permissible capability set. When PaymentAgent-Acquirer next attempts a high-value A2A payment authorization request to PaymentGateway-PSP (\url{a2a://Gateway.PaymentAuth.PSPGlobal.v4.secure::did:...}), the AEM at the gateway side consults the SSS. Even if the agent presents its usual VCs, the SSS indicates "ReducedTrust." The PDP at the gateway might now enforce a stricter policy: "IF agent\_status == 'ReducedTrust', THEN require\_multi\_factor\_agent\_auth (e.g., a ZKP of a recent controller approval for this transaction type) OR limit\_transaction\_value\_to\_low\_threshold." The A2A transaction might be rejected or queued for additional checks, preventing potential fraud by a slightly misbehaving or partially compromised agent.
\end{itemize}
The ANS provides discoverable context about an agent's intended role and capabilities. Monitoring deviations from this publicly or organizationally declared purpose, in addition to private policy violations, gives a richer signal for anomaly detection. The trust score becomes more robust as it can factor in the consistency of an agent's behavior with its registered identity profile in ANS.

\subsection{Agile Incident Response: Precision Targeting, Rapid Containment, and Discoverable Impact}

When a security incident occurs, the ability to respond swiftly, precisely, and comprehensively is critical to minimizing damage. The proposed IAM framework, with its integration of DIDs, VCs, and ANS, provides superior capabilities for incident response.

\textit{Rapid and Unambiguous Identification via DID and ANS Context:} Security alerts from monitoring systems or external threat intelligence will directly reference the compromised or malicious agent's DID and often its ANSName. This removes ambiguity and allows response teams to immediately identify: The specific agent instance involved (via DID). Its declared purpose and owner (via ANSName and resolved DID document). Its attested capabilities and dependencies (via VCs and DID document).

\textit{Example:} An alert "Unusual data exfiltration by did:com:cloudstorage:agent:backup-beta-721 (ANSName: \url{a2a://Backup.CriticalDB.AcmeCorp.v1.beta.nightly})" immediately tells the SOC: It's a specific backup agent instance. It's associated with AcmeCorp's critical database backups. It's a beta version (which might imply higher risk or different oversight).

\textit{Targeted Revocation with Ecosystem-Wide Propagation:} The framework supports granular to broad revocation, propagated efficiently:
\begin{itemize}
    \item \textbf{VC Revocation (Surgical):} If a specific attested capability (e.g., VC:AbilityToModifyUserPermissions) of AdminBot-HR (did:com:hr:adminbot:003, ANSName \url{a2a://UserAdmin.Permissions.HRInternal.v2.prod}) is found to be exploited due to a bug, that VC is added to a VC Status List. AdminBot-HR might still function for other tasks (e.g., reading user profiles) using its other VCs, but attempts to use the revoked permission VC will fail.
    \item \textbf{DID Deactivation/Revocation (Logical via DID Method or ANS):} If AdminBot-HR's private keys are confirmed stolen, its entire DID (did:com:hr:adminbot:003) is revoked via its DID method. The ANS entry for \url{a2a://UserAdmin.Permissions.HRInternal.v2.prod} would then either resolve to a "revoked" status or be removed/updated by the ANS Registration Authority. Other agents querying ANS for this service will no longer receive the compromised DID.
    \item \textbf{Instantaneous Global Session Invalidation via Unified Enforcement Layer:} This is the most critical response.
    \begin{itemize}
        \item Trigger: SOC confirms did:com:hr:adminbot:003 is actively malicious.
        \item SA Notification: The Session Authority (SA) is notified, specifying the DID.
        \item SSS Update: SA updates the Session State Synchronizer (SSS) to mark all global sessions for did:com:hr:adminbot:003 as "TERMINATED\_IMMEDIATE\_SECURITY\_LOCKOUT".
        \item AEM Enforcement: All AEMs interacting with or receiving requests from did:com:hr:adminbot:003 (whether via A2A, MCP, or internal ACP/HTTP calls) consult the SSS. They see the "TERMINATED" status and instantly block any new requests and terminate any active local protocol sessions.
        
        \textit{Scenario (MCP Tool in use by AdminBot-HR):} If AdminBot-HR was using an MCP tool like UserProvisioningTool, its active MCP session (managed by the tool's AEM) would be killed. Further MCP calls from AdminBot-HR would be rejected by the AEM before even reaching the tool's logic.
        
        \textit{Scenario (A2A communication):} If AdminBot-HR was sending A2A messages to AuditLogAgent, these A2A messages would be blocked by the AEM on AuditLogAgent's side.
    \end{itemize}
\end{itemize}
The ANS provides a clear point for signaling revocation at the discovery layer. Even if an attacker has cached an old DID, new discovery attempts for the agent's function would fail or return a revoked status. The SSS ensures that active sessions, regardless of how they were initiated (perhaps post-ANS discovery), are comprehensively terminated.

\textit{Rich Forensic Analysis with Discoverable Context:} Post-incident, the combination of DID-anchored logs, VCs, and ANS information provides unparalleled depth for forensics.
\begin{itemize}
    \item \textbf{Contextualizing Compromise:} If did:com:research:agent:dataminer:gamma-9 is compromised, investigators can not only see its actions (via DID logs) but also: Resolve its ANSName (\url{science://DataMining.LargeDatasets.ResearchDiv.v0.9.experimental}) to understand its expected role and provider context. Examine its DID Document and VCs to see its intended capabilities and dependencies (e.g., "depends on did:com:lib:math:vectorcalc:v3.2"). This helps to check if a dependency was the root cause. Trace its ANS query history: Was it trying to discover and interact with services outside its normal profile before the compromise? If it interacted with other agents, their DIDs/ANSNames are in the logs, allowing investigators to assess the blast radius and check if those collaborators were also affected or were part of the attack.
    \item \textbf{Identifying Attack Vectors via ANS:} If multiple agents registered under a specific, less reputable Provider in their ANSNames are simultaneously compromised, it might indicate a targeted attack against that provider's agent infrastructure or a vulnerability common to their agents.
\end{itemize}
ANS data (like provider, capability domain in the name) adds valuable metadata for clustering incidents, identifying patterns, and understanding the potential scope or origin of an attack that might involve multiple agent instances from a similar source or with similar functions.

\subsection{Other Potential Uses Building on Verifiable Agent IDs and Discoverable ANS Profiles}

The synergistic use of detailed, verifiable Agent IDs and a structured Agent Name Service, all managed within a robust IAM framework, naturally extends to enable further advanced functionalities critical for a mature and trustworthy AI ecosystem.

\textit{Decentralized Reputation and Trust Brokering with ANS-Contextualized Feedback:}
\begin{itemize}
    \item Agent DIDs serve as the stable anchors for accumulating reputation scores. When AgentA (e.g., discovered via ANS as \url{a2a://TaskExecutor.GeneralPurpose.CommunityPool.v1.standard::did:agentA...}) completes a task for AgentB, AgentB can issue a reputation VC attesting to AgentA's performance, timeliness, and reliability for that specific task type (derived from AgentA's ANS capability).
    \item These VCs can be stored by AgentA or published to a decentralized reputation ledger. Future agents querying ANS for "TaskExecutor.GeneralPurpose" might then also be able to query this reputation system (using the resolved DID) for community feedback, prioritizing agents with higher, relevant reputation scores. The ANS capability string itself provides context for the reputation (e.g., good at "GeneralPurpose" tasks).
\end{itemize}
Code Concept: Agent B issuing a reputation VC for Agent A:
\begin{lstlisting}[style=customcode, caption=Agent B Issuing Reputation VC for Agent A, label=lst:reputation_vc_code, language=Python]
# Agent B's perspective
# from pyld import jsonld # For Verifiable Credentials
# from did_sdk import sign_vc # Conceptual SDK function

agent_A_did = "did:agentA..." 
agent_A_ans_capability = "TaskExecutor.GeneralPurpose."
    + "CommunityPool.v1.standard"

reputation_claim = {
    "@context": ["https://www.w3.org/2018/credentials/v1", 
                 "https://example.org/reputation/v1"],
    "type": ["VerifiableCredential", "ReputationCredential", "PerformanceReview"],
    "issuer": "did:agentB...", # Agent B's DID
    "issuanceDate": "2025-10-03T10:00:00Z",
    "credentialSubject": {
        "id": agent_A_did,
        "ansCapabilityContext": agent_A_ans_capability,
        "rating": 5, # Scale of 1-5
        "comment": "Completed task efficiently and accurately.",
        "taskId": "task-uuid-for-context"
    }
}

# Agent B signs this claim with its DID key to create a VC
# signed_reputation_vc = sign_vc(reputation_claim, "did:agentB...", "did:agentB...#key-1")

# Agent B might then send this VC to Agent A, or publish it to a reputation service.
\end{lstlisting}

\textit{Automated Billing and Resource Quota Enforcement via ANS-Defined Services:}
\begin{itemize}
    \item When an agent discovers and uses a commercial service (e.g., a specialized MCP tool like \url{mcp://AdvancedTranslation.Multilingual.PremiumAPI.v3.commercial::did:tool:translateXYZ...}) via ANS, the ANS record itself might point to metadata about pricing models or rate limits associated with that service DID.
    \item The consuming agent's DID is logged by the commercial tool for every API call. The tool provider's AEM/PDP can enforce quotas (e.g., "Agent did:com:startup:agent:translator007 has a quota of 10M characters/month for did:tool:translateXYZ"). Billing is then accurately attributed to the consuming agent's controller.
\end{itemize}

\textit{Secure Software/Model Supply Chain Attestations Linked to ANS Registrations:}
\begin{itemize}
    \item When an agent is registered with ANS (e.g., \url{a2a://ImageRecognition.MedicalScans.RadAI.v2.validated::did:radai:imgrec:002}), part of its registration with the ANS Registration Authority (RA) could involve presenting VCs that attest to its supply chain security: A VC for its base foundation model (e.g., "ModelCard\_VC\_for\_RadAI\_BaseVisionModel\_v2"), detailing its training data, bias tests, and safety evaluations. SBOM VCs for its software components. A "ValidatedSecureBuild\_VC" from a trusted CI/CD pipeline.
    \item The ANS resolver could then optionally return indicators of these attestations (or links to the VCs) along with the agent's DID, allowing discoverers to prioritize agents with verifiable supply chain security.
\end{itemize}

\textit{Dynamic Coalition Formation and Capability Negotiation using ANS for Initial Matching:}
\begin{itemize}
    \item An EmergencyResponseOrchestratorAgent queries ANS for agents with diverse capabilities like \url{a2a://DroneSurveillance.DisasterZoneMapping...}, \url{mcp://Logistics.ResourceAllocation...}, and \url{comms://TemporaryNetwork.MeshDeployment....}
    \item Once candidate DIDs are retrieved, the orchestrator can initiate a negotiation phase (e.g., using FIPA Contract Net Protocol~\cite{FIPA2002ContractNet} messages over A2A or ACP). During negotiation, agents exchange more detailed VCs about their specific sub-capabilities, current availability, and resource needs.
    \item The orchestrator then issues a "CoalitionCharter\_VC" to the selected agents, defining the coalition's DID, its mission, shared resources (perhaps managed by a temporary group DID), roles, and duration. This VC acts as a temporary authorization within the coalition.
\end{itemize}

\textit{ANS for Discovering Ethical AI Governance Services:}
\begin{itemize}
    \item Agents or users could query ANS for services like \url{audit://EthicalComplianceOracle.AIBehavior.IndependentOrg.v1} or \url{report://BiasReportingService.FairnessConsortium.v1}.
    \item These specialized services (themselves having DIDs and VCs) could then be used by agents to self-assess their decisions against ethical guidelines or for users to report problematic agent behavior, with the AN DIDs providing a verifiable link to the service.
\end{itemize}
By integrating ANS as a core discovery mechanism whose results (DIDs, initial capability claims) feed directly into the DID/VC-based authentication and authorization processes, the entire IAM lifecycle becomes more context-aware, secure, and efficient. The discoverable nature of agent capabilities and attestations fosters a more transparent and trustworthy ecosystem.

\section{Deployment Models \& Governance Considerations}

The proposed Agentic AI IAM framework, while architecturally comprehensive, is not a monolithic, one-size-fits-all solution in terms of its practical implementation. The diverse needs of different organizations, Multi-Agent System (MAS) scopes (private enterprise vs. open ecosystem), trust requirements, and existing infrastructure will necessitate different deployment models for its core components (e.g., DID registries, Verifiable Credential (VC) issuers, Agent Name Service (ANS), Policy Engines, Session Authority, Session State Synchronizer). Furthermore, regardless of the chosen deployment model, robust, well-defined, and adaptable governance is paramount for the long-term viability, trustworthiness, security, and interoperability of any such advanced IAM system.

\subsection{Deployment Model Analysis}

We analyze three primary deployment models—Centralized, Decentralized, and Federated—assessing their characteristics, advantages, disadvantages, and suitability for various Agentic AI IAM scenarios.

\subsubsection{Centralized Approach}
\textit{Description:} In a centralized deployment, a single organization, platform provider, or a designated administrative entity controls and operates all, or the significant majority, of the IAM framework's core components. This typically includes: The primary Agent ID registry (which might be a private Public Key Infrastructure (PKI) issuing X.509 certificates as per some ANS proposals, a proprietary database issuing unique identifiers, or a private DID method controlled by the organization). The authoritative VC issuers for organizational roles, capabilities, and compliance attestations. The ANS, if implemented as a private or enterprise-scoped directory service. The central Policy Decision Points (PDPs) and Policy Administration Points (PAPs) defining and enforcing access rules. The Cross-Protocol Session Authority (SA) and the Session State Synchronizer (SSS). Agents operating within this model typically belong to, or are tightly managed and permissioned by, the central entity. All trust decisions ultimately flow from this central authority.

\textit{Advantages:} Simplified Governance \& Policy Cohesion. Unified Control, Visibility, and Audit. Potentially Easier Integration with Existing Enterprise Systems. Optimized Performance. Clear Accountability.

\textit{Disadvantages:} Single Point of Failure, Control, and Trust. Scalability Bottlenecks. Vendor/Platform Lock-in. Limited Cross-Organizational Trust \& Interoperability. Potential for Censorship or Abuse of Power.

\textit{When to Use:} Enterprise-Internal MAS. Specific AI Platforms. Early-Stage Deployments or Controlled Experiments. Highly Regulated Environments with a Single Auditing Authority.

\subsubsection{Decentralized Approach}
\textit{Description:} Core IAM components are implemented using decentralized technologies, often public and permissionless, or permissioned consortia-based Distributed Ledger Technologies (DLTs). Key characteristics include: DIDs are registered on public or consortia DLTs (e.g., did:ion, did:ethr, did:sov, or a custom agent-focused DID method on a dedicated ledger like the proposed Agent ID Provider Network - AIPN). Agent controllers or agents themselves manage their DID's private keys. VCs can be issued by a diverse set of issuers (each with their own DID) and their status (revocation) might be tracked via decentralized mechanisms (e.g., on-chain registries, distributed VC status lists). ZKPs are used extensively for privacy-preserving presentation of VCs and attributes. ANS could be built on decentralized name systems (e.g., ENS, Handshake, or a custom DLT-based ANS). Policy enforcement might involve smart contracts acting as rudimentary PDPs for on-chain resources, or rely on Verifiable Presentations that bundle VCs required by a verifier's policy. Global session state (like revocation lists) might be mirrored on resilient DLTs. Governance is typically community-driven (e.g., DAOs for protocol upgrades) or based on the immutable logic encoded in smart contracts.

\textit{Advantages:} No Single Point of Failure or Control. User/Agent Sovereignty (SSI). Enhanced Trust in Open, Permissionless Ecosystems. Transparency \& Auditability (for public DLTs). Censorship Resistance.

\textit{Disadvantages:} Governance Complexity, and "Tragedy of the Commons". Smart Contract and DLT Security Risks. Performance, Scalability, and Cost of DLTs. User/Controller Experience (Key Management). Irreversibility and Data Privacy. Bootstrapping Trust. 

\textit{When to Use:} Truly Open, Permissionless Multi-Agent Ecosystems. Cross-Organizational Collaborations Without a Central Trusted Party. Applications Requiring Very High Degrees of Censorship Resistance or User Control Over Identity. Ecosystems Where a Transparent, Community-Governed Trust Infrastructure is a Core Design Goal.

\subsubsection{Federated Approach}
\textit{Description:} This model involves multiple independent IAM domains or "trust communities." Each domain might manage its own IAM infrastructure using centralized or even localized decentralized approaches. The key is that these domains establish mutual trust relationships and define standardized protocols for interoperability. This could involve: Cross-certification of Certificate Authorities (CAs) or DID method roots between domains. Shared trust lists for recognized VC issuers and verifier policies across the federation. Federated ANS resolution (e.g., similar to how DNS subdomains can be delegated, or using inter-registry lookup protocols). Use of highly interoperable DID methods and standardized VC profiles (e.g., based on W3C specs) to ensure credentials from one domain can be understood and verified in another. A central (or mutually agreed upon) body might define the "federation rules" or baseline interoperability standards, but day-to-day IAM within each domain remains autonomous.

\textit{Advantages:} Balances Autonomy with Interoperability. Scalability. Domain-Specific Policies and Trust Levels. Enhanced Resilience. Phased Adoption \& Existing System Integration.

\textit{Disadvantages:} Complexity of Trust Management. Interoperability Challenges (Technical and Semantic). Potential for Lowest Common Denominator Security. Discovery and Pathfinding Complexity. Governance Overhead for the Federation Itself.

\textit{When to Use:} Consortia of Organizations in a Specific Industry. Alliances of Research Institutions or Governmental Agencies. Large, Multi-National Corporations with Distinct Regional or Business Unit IAM Requirements. Ecosystems Evolving from Existing Centralized or Siloed Systems Towards Greater Interoperability. As a practical model for the Agent Name Service (ANS).

\subsubsection{Hybrid Approaches}
It's important to note that these models are not always mutually exclusive. Hybrid approaches are likely to be common, combining elements from each.
\begin{itemize}
    \item \textit{Example 1:} An enterprise might use a centralized IAM framework for its internal agents but use a federated model to interact with agents from trusted partners. Its internal agents might have DIDs issued by a private DID method, but these DIDs could be anchored or discoverable through a broader federated system.
    \item \textit{Example 2:} A decentralized ecosystem might still rely on a few, highly reputable (perhaps foundation-run) "anchor" VC issuers for certain critical credentials (like "VerifiedLegalEntity\_VC"), even if most other VCs are issued more peer-to-peer.
    \item \textit{Example 3:} The Session Authority and Session State Synchronizer, while logically providing global coordination, might be implemented as a permissioned DLT operated by a consortium (federated control over a logically centralized function) for resilience and shared trust.
\end{itemize}

\subsection{Decision Matrix for Choosing an Implementation Model}
Selecting the most appropriate deployment model requires careful consideration of various factors. Table~\ref{table_decision_matrix} provides guidance:

\begin{table*}[!t]
\renewcommand{\arraystretch}{1.5} 
\caption{Decision Matrix for Choosing an Implementation Model}
\label{table_decision_matrix}
\centering
\scriptsize 
\begin{tabularx}{\textwidth}{@{}l >{\raggedright\arraybackslash}X >{\raggedright\arraybackslash}X >{\raggedright\arraybackslash}X >{\raggedright\arraybackslash}X@{}} 
\toprule
\textbf{Feature / Requirement} & \textbf{Centralized} & \textbf{Decentralized} & \textbf{Federated} & \textbf{Hybrid} \\
\midrule
Control Authority & Single Entity (High Control) & Community/Protocol (Low Central Control) & Domain-Specific + Federation Body (Balanced) & Varies; often domain-specific with shared elements \\ 
Trust Model & Hierarchical (Trust in Central Entity) & Peer-to-Peer / Ledger-Based (Distributed Trust) & Inter-Domain Agreements / Shared Roots (Delegated) & Mix of hierarchical and delegated/distributed \\
Scalability & Moderate (Potential Bottlenecks) & Potentially Very High (if DLT scales) / Variable & High (Distributed across domains) & High (Can optimize components) \\
Performance (Latency) & Potentially Low (if optimized, local) & Variable (DLT dependent, often higher) & Moderate (Inter-domain hops) & Variable (Can optimize critical paths) \\
Interoperability (External) & Low (Proprietary by default) & Potentially High (if open standards used) & High (Designed for inter-domain ops) & Moderate to High (Depends on bridge design) \\
Complexity of Setup & Low to Moderate & High & High (Trust agreements complex) & Moderate to High \\
Complexity of Governance & Low (Single decision-maker) & Very High (Consensus, community) & High (Federation rules, inter-domain) & High (Managing diverse components) \\
Cost (Infrastructure) & Moderate (Centralized infra) & Variable (DLT fees can be high) & Moderate to High (Per-domain + federation infra) & Variable \\
Security (vs External Threats) & Single attack surface (high impact if breached) & Distributed risk, smart contract vulns critical & Risk shared/isolated per domain; inter-domain trust & Tailorable; can have strong internal, defined external \\
User/Agent Sovereignty & Low & Very High & Moderate (Within domain policies) & Variable \\
Censorship Resistance & Low & High & Moderate (Per domain) & Variable \\
Privacy Preservation & Dependent on central entity's policies & High (with ZKPs, careful DLT use) & Domain-specific policies; inter-domain data flow & Can be designed for high privacy \\
Suitability: Enterprise Internal & High & Low & Moderate (For large, distinct internal units) & High (Central core, federated edges) \\
Suitability: Open Ecosystem & Low & High & Moderate (Federation of open communities) & Moderate (Public services with private backends) \\
Suitability: B2B Consortia & Low (Unless one org dominates) & Moderate (If common DLT agreed) & High & High (Federated interfaces, shared services) \\
\bottomrule
\end{tabularx}
\end{table*}

\textit{How to use the matrix:}
\begin{enumerate}
    \item Identify the primary context for your MAS (e.g., internal enterprise, open research platform, industry consortium).
    \item Prioritize your key requirements (e.g., is maximum agent sovereignty critical, or is centralized auditability paramount?).
    \item Evaluate each model against your high-priority requirements.
    \item Consider if a hybrid approach offers the best trade-offs by combining strengths of different models for different IAM components (e.g., decentralized DIDs but a federated or even centrally managed Session Authority for specific use cases).
\end{enumerate}

\subsection{Governance Considerations}
Effective governance is the bedrock upon which trust and interoperability in any Agentic AI IAM framework are built. It's not merely about technical rules but also about establishing clear roles, responsibilities, processes for decision-making, dispute resolution, and adaptation over time.

\subsubsection{Identity Governance (DIDs, VCs, ANS)}
\begin{itemize}
    \item DID Method Governance
    \item ANS Namespace Management \& Policy
    \item VC Issuer Accreditation, Trust Registries, and Governance Frameworks
    \item Agent ID Lifecycle Management Policies
\end{itemize}

\subsubsection{Security Policy Governance (for PDPs and SA)}
\begin{itemize}
    \item Policy Authorship \& Approval Workflows
    \item Policy-as-Code Principles
    \item Emergency Policy Override Procedures
    \item Policy Interoperability/Harmonization (in Federated Models)
\end{itemize}

\subsubsection{Operational and Security Governance for IAM Infrastructure}
\begin{itemize}
    \item Incident Response Playbooks for IAM Breaches
    \item Key Management Governance for IAM Services
    \item Regular Audits \& Penetration Testing
    \item Vulnerability Disclosure Policy
\end{itemize}

\subsubsection{Data Privacy and Ethical Use Governance}
\begin{itemize}
    \item Data Protection Impact Assessments (DPIAs) for IAM Data
    \item Agent ID Data Minimization Principles
    \item Bias Review in Credentialing and Reputation
    \item Ethical Oversight Bodies
\end{itemize}

\subsubsection{Evolution and Standards Governance}
\begin{itemize}
    \item Change Management Process
    \item Liaison with External Standards Bodies
\end{itemize}
Effective governance in the Agentic AI IAM space will not be static; it must be an adaptive system capable of evolving alongside the technology and the threat landscape. It necessitates a collaborative effort, potentially involving a mix of industry self-regulation, standards development, and, where appropriate, governmental oversight, particularly for public-facing or critical infrastructure components.

\section{Security Considerations}

Securing the Agentic AI IAM framework is paramount, analyzed here using the MAESTRO framework \cite{Huang2025bMAESTRO}.

\subsection{The MAESTRO 7-Layer Reference Architecture for Agentic AI}
MAESTRO decomposes AI ecosystems into: Layer 1: Foundation Models, Layer 2: Data Operations, Layer 3: Agent Frameworks, Layer 4: Deployment and Infrastructure, Layer 5: Evaluation and Observability, Layer 6: Security and Compliance (Vertical), and Layer 7: Agent Ecosystem.

\subsection{Threat Analysis of the Proposed Agentic AI IAM Framework using MAESTRO Layers}
\begin{itemize}
    \item \textbf{L1: Foundation Models:} Model-based identity theft that occurs when attackers use AI models to analyze and replicate the behavioral patterns, communication styles, and decision-making characteristics of legitimate agents, effectively creating digital impersonators that can fool other systems or users into believing they're interacting with the authentic agent (mitigated by cryptographic DIDs/VCs).
    \item \textbf{L2: Data Operations:} Poisoning of DID registries/VC status lists (mitigated by DLT consensus, signed registry entries); exfiltration of identity data (mitigated by encryption, agent-held VCs, ZKPs); tampering with PIPs (mitigated by PIP identity and secure channels).
    \item \textbf{L3: Agent Frameworks:} Compromised IAM SDKs (mitigated by secure development, sandboxing); framework vulnerabilities allowing session hijacking (mitigated by continuous re-validation via AEM/SSS).
    \item \textbf{L4: Deployment and Infrastructure:} DoS/DDoS against IAM services (mitigated by standard defenses, resilient design); compromise of IAM service infrastructure (mitigated by hardening, access controls); lateral movement to IAM components (mitigated by network segmentation, Zero Trust).
    \item \textbf{L5: Evaluation and Observability:} Tampering with IAM audit logs (mitigated by immutable logging, signatures); evasion of IAM monitoring (mitigated by comprehensive instrumentation); data leakage via observability tools (mitigated by masking, ZKPs).
    \item \textbf{L6: Security and Compliance:} Misconfiguration of IAM policies (mitigated by policy-as-code, audits); compromise of IAM service keys (mitigated by HSMs, revocation); non-compliance with privacy regulations (mitigated by privacy-by-design, ZKPs).
    \item \textbf{L7: Agent Ecosystem:} Agent impersonation/DID spoofing (mitigated by cryptographic verification, VC status checks); compromised ANS leading to malicious discovery (mitigated by secure ANS resolution); collusion to falsify VCs (mitigated by trust diversification, reputation systems).
\end{itemize}

\subsection{Cross-Layer Threats Affecting the IAM Framework}
Including supply chain attacks on IAM components, privilege escalation across IAM layers, and goal misalignment leading to IAM misuse, all requiring defense-in-depth and continuous monitoring.

\subsection{Applying Zero Trust to Agentic AI IAM Framework}
This brings essential security, governance, and accountability benefits especially given the autonomous decision making, undeterministic behavior, and scale of AI agents. Implemented and tested security controls that are preventative, detective, and corrective form the basis of Zero Trust. These fundamentals are critical to the success of a Zero Trust implementation: Concept of least-privilege access, Separation of duties, Segmentation/micro-segmentation, Logging and monitoring, Configuration drift remediation, Assume breach, Dynamic and adaptive security policy enforcement.

\section{Innovative Contributions of this Framework}

The proposed framework represents a significant departure from traditional approaches, offering a collection of synergistic innovations specifically designed for the unique challenges of autonomous Multi-Agent Systems (MAS). These contributions are not isolated features but form part of a re-conceptualization of agent identity, integrating advanced cryptographic techniques and a novel architectural design for dynamic control, all within a holistic, lifecycle-aware approach to managing AI agents as first-class digital citizens.

The foremost contribution is the articulation of a comprehensive, end-to-end IAM framework purpose-built for the agentic paradigm. This moves beyond merely adapting human-centric or simplistic machine and NHI (Non Human Identity) IAM protocols, which often prove inadequate for the complexities of autonomous, interacting agent swarms. Instead, our framework cohesively integrates identity issuance, rich credentialing, capability-aware discovery, dynamic access control, and a novel cross-protocol enforcement layer into a unified conceptual model. It addresses the entire lifecycle of an agent—from its "birth" through its operational interactions to its eventual decommissioning—recognizing the deep interdependencies between these stages. Existing IAM solutions typically focus on narrower problems, struggling with identities that spawn others, dynamically change roles, or require fine-grained, context-sensitive authorization at massive scale. This framework's systemic integration is designed to address these fundamental gaps.

Central to this is a redefinition of Agent Identity, making it rich, dynamic, and verifiably secure. We shift away from simplistic identifiers like API keys towards identities anchored by cryptographically secure Decentralized Identifiers (DIDs). This DID-anchored identity is not static; it is an extensible digital representation augmented by Verifiable Credentials (VCs) that attest to an agent's attributes, capabilities, compliance status, roles, and provenance. The dynamism is crucial, as AI agents evolve, their models update, capabilities expand, and compliance needs re-attestation. A rich, verifiable identity containing fields like scopeOfBehavior, toolset (which can include DIDs of authorized tools), modelHash, and VCs for training data or compliance, allows for far more nuanced trust and authorization. The use of DIDs provides self-sovereignty and interoperability, essential for open MAS, while VCs offer a standardized, vendor-neutral way to make diverse claims. Furthermore, Zero-Knowledge Proofs (ZKPs) enable agents to selectively and privately present these verifiable claims, a significant advancement over the limited flexibility and privacy of traditional certificate extensions.

Building on this rich identity, the framework introduces capability-centric discovery and more granular access control. An integrated Agent Naming Service (ANS) facilitates secure discovery, allowing agents to find others not just by name but by the specific functions or attested capabilities they offer. This is a critical distinction from traditional service discovery, which may locate an endpoint but doesn't inherently verify the target’s attested abilities. Our approach directly links discovery to verifiable identity attributes. Authorization decisions thereby become more intelligent, considering not just "who" is making a request, but fundamentally "what is this agent verifiably capable and authorized to do, with which specific tools, and under what attested conditions?" By making an agent’s authorized toolset and scopeOfBehavior verifiable parts of its identity, the system can enforce the principle of least function, significantly limiting the blast radius of a compromised or misbehaving agent. The framework introduces Context-Based Access Control which enables dynamic access decisions based on real-time environmental, behavioral, and task context moving beyond static roles or attributes allowing enforcement policies to adapt to an agent’s current state and conditions.

A cornerstone innovation is the Unified Cross-Protocol Global Session Management and Policy Enforcement Architecture. This Layer 4 uniquely addresses the challenge of maintaining consistent security posture in heterogeneous MAS where agents use diverse communication protocols. In such environments, a critical security gap is the inability to propagate vital IAM state changes—like a global session termination, a master DID revocation, or a sudden capability downgrade—instantaneously and uniformly across all interaction points. This layer acts as a "security and session management backplane," ensuring that a policy decision or revocation, once made, is effectively and immediately enforced wherever an agent might interact, regardless of the underlying transport. This real-time, cross-protocol consistency is fundamental for operationalizing robust security.

The framework also achieves a pragmatic fusion of self-sovereignty with enforceable governance. While DIDs and agent-controlled VCs empower agents and their controllers with greater control over their core identity data, this self-sovereignty is balanced with mechanisms for practical governance. This means that while an agent can present its self-managed identity, these credentials can be verified against established trust frameworks, such as lists of accredited VC issuers for specific roles or compliance attestations. The Session Authority retains the ability to enforce global revocations or policy overrides based on enterprise risk decisions, even if the agent "controls" its DID. This balance is vital for adoption in real-world systems that require clear lines of accountability and cannot operate solely on peer-to-peer trust.

Finally, the framework provides intrinsic support for fine-grained accountability and verifiable provenance. Cryptographic verifiability is embedded at multiple levels: for identities via DIDs and their keys; for claims about agents via VCs and issuer signatures; for agent actions using the agent's DID-associated private key. The Agent ID structure itself is designed to encapsulate or link to detailed provenance information—such as its creator, constituent models, software dependencies (potentially with their own DIDs), and VCs attesting to training data or safety audits. As AI agents are entrusted with increasingly impactful decisions, the ability to irrefutably determine "who (which agent instance) did what, when, why, with what authority, and based on what information/capabilities" becomes critical. This moves beyond basic logging to establish a cryptographically verifiable audit trail, essential for forensics, dispute resolution, and building societal trust in autonomous systems, directly addressing the "audit trail ambiguity" prevalent in current systems and providing a much stronger basis for non-repudiation.

To measure the success implementation of the innovation, the following Key Performance Indicators (KPIs) can be considered: Successful Agent Authentication Rate, Authorization Latency, Policy Enforcement Accuracy, Revocation Time, Audit Log Integrity, Anomaly Detection Rate, Incident Response Time, Agent Discovery Success Rate, Downtime due to IAM Issues.

\section{Discussion and Future Work}

As future work, we have identified the following.

\subsection{Scalability, Performance, and Efficiency}
The Challenge: Several components within the proposed architecture, particularly those involving Distributed Ledger Technologies (DLTs) for DID registration and Verifiable Credential (VC) status management, or the Session State Synchronizer (SSS) which must track potentially millions of active agent sessions, face significant scalability and performance hurdles. The cryptographic operations inherent in DIDs, VCs, and Zero-Knowledge Proofs (ZKPs), while providing security, can also introduce computational overhead for resource-constrained agents or high-throughput systems.

Future Work: Benchmarking and Optimization; Efficient Cryptography; Caching and Resolution Strategies; Hardware Acceleration.

\subsection{Standardization and Interoperability}
The Challenge: The true power of a global Agentic AI IAM framework lies in its interoperability. Without widely adopted standards for how Agent IDs are structured, how capabilities are defined and attested in VCs, how ZKPs are constructed for common proofs, or how ANS queries are formatted, the ecosystem risks fragmentation into incompatible identity silos.

Future Work: Active Standards Development; Agent-Specific Profiles; Common Ontologies; Reference Implementations and Conformance Suites; Formalization of Model Context Protocols (MCPs).

\subsection{Governance Models, Trust Frameworks, and Legal Considerations}
The Challenge: Establishing and managing governance for a potentially global, decentralized, or federated IAM infrastructure is a monumental task. This includes defining who can issue authoritative VCs (e.g., for legal identity or compliance), how disputes over DIDs or ANS names are resolved, and how liability is attributed in complex MAS interactions. The evolving legal and regulatory landscape for AI also presents a moving target.

Future Work: Multi-Stakeholder Governance Research; Trust Assurance Levels; Legal and Regulatory Analysis; Dispute Resolution Mechanisms; Security Controls Specific to AI Agents.

\subsection{Enhanced Security and Privacy in Practice}
The Challenge: While the framework incorporates strong security primitives, sophisticated adversaries will inevitably seek to exploit implementation weaknesses, social engineering aspects, or unforeseen interaction effects between components. Maintaining agent and user privacy in the face of increasingly rich identity data is also paramount.

Future Work: Formal Security Modeling and Verification; Agent-Specific Threat Intelligence; Advanced Privacy-Enhancing Technologies (PETs); Secure Key Management for Autonomous Agents; Resilience Against Quantum Threats; Tabletop Exercises for Agentic Incident Response.

\subsection{User Experience (UX), Developer Tooling, and Adoption Pathways}
The Challenge: For this framework to be adopted, it must be usable by both end-users (who may act as controllers for their personal agents) and developers building and deploying AI agents. Complexity in managing DIDs, VCs, and policies can be a significant barrier.

Future Work: Developer-Friendly SDKs and Libraries; Management UIs and Dashboards; "Secure by Default" Agent Architectures; Phased Adoption Strategies.

\subsection{Ethical Considerations and Societal Impact Mitigation}
The Challenge: The power of verifiable and persistent Agent IDs, while beneficial for security, also carries potential risks if misused for pervasive surveillance, biased decision-making (e.g., if VCs for "good behavior" are only available to certain types of agents), or creating new forms of digital divide.

Future Work: Ethical Impact Assessments; Bias Detection and Mitigation in Credentialing; Transparency and Explainability of IAM Decisions; Public Discourse and Inclusive Design.

The journey to a fully realized and globally functional Agentic AI IAM framework is an ambitious one. It necessitates a collaborative, iterative approach, blending cutting-edge research with pragmatic engineering and a deep understanding of the evolving societal context of AI. Addressing these future work areas will be critical to transforming the vision presented in this paper into a resilient, trustworthy, and enabling infrastructure for the future of AI.

\section*{Acknowledgment}
The authors extend their deepest gratitude to the following individuals whose expertise, insights, and collaborative spirit made this research on Agentic AI Identity and Access Management approaches possible.

\textit{Authors and Contributors (implicitly covered by the author list, but acknowledged as per original text):} Vineeth Sai Narajala, John Yeoh, Json Ross, Mahesh Lambe, Ramesh Raskar, Youssef Harkati, Jerry Huang, Chris Hughes.

\textit{Reviewer:} Idan Habler, PhD, Staff AI/ML Security Researcher at Intuit, whose thorough review and constructive feedback significantly improved the quality and accuracy of this research.

\textit{Brainstorming Contributors:} Special appreciation goes to the following thought leaders who have discussed with Ken Huang on Agentic AI security on many different occasions in his involvement as co-chair of the CSA AI Safety Working groups, at the RSA 2025 conferences, and LinkedIn discussions: Jim Reavis, CEO and Founder of Cloud Security Alliance; Daniele Catteddu, Chief Technology Officer at Cloud Security Alliance; Caleb Sima, Chair of CSA AI Security Alliance; Professor Dawn Song of the University of California, Berkeley; Michael Bargury, Co-founder and CTO of Zenity; Dr. Chenxi Wang of Rain Capital; Nate Lee, Founder of Cloudsec.ai; Ed Sewell, NVIDIA AI INCEPTION Member; Jojo Ye of Sixty Degree Capital; Akram Sheriff of Cisco.

This research on Agentic AI Identity and Access Management stands on the shoulders of these remarkable individuals, whose collective wisdom, diverse perspectives, and unwavering support made this work possible. Their contributions reflect the collaborative spirit essential to advancing Agentic AI Security.

\bibliographystyle{IEEEtran}
\bibliography{references}

\end{document}